\documentclass[12pt]{article}

\usepackage[english]{babel}
\usepackage[T1]{fontenc}

\usepackage[a4paper,top=2cm,bottom=2cm,left=2cm,right=2cm,marginparwidth=1.75cm]{geometry}

\usepackage{amsmath, amsthm, amssymb}
\usepackage{graphicx}
\usepackage[natbib, maxcitenames=5, style=apa, uniquename=false, uniquelist=false]{biblatex}
\usepackage{csquotes}
\numberwithin{table}{section}

\usepackage{setspace}

\usepackage[colorlinks=true, allcolors=blue,hypertexnames=false]{hyperref}

\newtheorem{tw}{Theorem}
\newtheorem{remark}{Remark}

\newcommand{\bX}{\boldsymbol{X}}
\newcommand{\bx}{\boldsymbol{x}}

\newcommand{\by}{\boldsymbol{y}}
\newcommand{\bh}{\boldsymbol{h}}

\newcommand{\ba}{\boldsymbol{a}}

\newcommand{\bw}{\boldsymbol{w}}
\newcommand{\bd}{\boldsymbol{d}}

\newcommand{\bz}{\boldsymbol{z}}
\newcommand{\bv}{\boldsymbol{v}}

\newcommand{\bU}{\boldsymbol{U}}
\newcommand{\bQ}{\boldsymbol{Q}}
\newcommand{\bG}{\boldsymbol{G}}

\newcommand{\bbeta}{\boldsymbol{\beta}}

\newcommand{\bgamma}{\boldsymbol{\gamma}}

\newcommand{\bPhi}{\boldsymbol{\Phi}}
\newcommand{\bEta}{\boldsymbol{\eta}}

\newcommand{\bZero}{\boldsymbol{0}}

\newcommand{\logit}{\operatorname{logit}}
\newcommand{\Exp}{\operatorname{Exp}}

\newcommand{\Uni}{\operatorname{Uniform}}
\newcommand{\argmin}{\operatorname{argmin}}

\title{Quantile balancing inverse probability weighting for non-probability samples}

\author{Maciej Beręsewicz\footnote{Poznań University of Economics and Business, Institute of Informatics and Quantitative Economics, Department of Statistics, Al. Niepodległości 10, 61-875 Poznań, Poland, E-mail: \url{maciej.beresewicz@ue.poznan.pl}; Statistical Office in Poznań, ul. Wojska Polskiego 27/29, 60-624 Poznań, Poland.}, Marcin Szymkowiak\footnote{Poznań University of Economics and Business, Institute of Informatics and Quantitative Economics, Department of Statistics, Al. Niepodległości 10, 61-875 Poznań, Poland; Statistical Office in Poznań, ul. Wojska Polskiego 27/29, 60-624 Poznań, Poland.}, Piotr Chlebicki\footnote{Stockholm University, Department of Mathematics, Albano hus 1, 106 91 Stockholm.}}
\date{}

\begin{document}
\maketitle

\doublespacing

\begin{abstract}
The use of non-probability data sources for statistical purposes and for official statistics has become increasingly popular in recent years. However, statistical inference based on non-probability samples is made more difficult by nature of their biasedness and lack of representativity. In this paper we propose \textit{quantile balancing inverse probability weighting estimator} (QBIPW) for non-probability samples. We apply the idea of Harms and Duchesne (2006) allowing the use of quantile information in the estimation process to reproduce known totals and the distribution of auxiliary variables. We~discuss the estimation of the QBIPW probabilities and its variance. Our simulation study has demonstrated that the proposed estimators are robust against model mis-specification and, as a~result, help to reduce bias and mean squared error. Finally, we applied the proposed methods to estimate the share of job vacancies aimed at Ukrainian workers in Poland using an integrated set of administrative and survey data about job vacancies.
\end{abstract}

\textbf{Key Words:} data integration, calibration approach, job vacancy survey.
\clearpage

\doublespacing

\section{Introduction}

In official statistics, information about the target population and its characteristics is mainly collected through probability surveys, census or is obtained from administrative registers, which cover all (or nearly all) units of the population. However, owing to increasing non-response rates, particularly unit non-response and non-contact resulting from the growing respondent burden, as well as rising costs of surveys conducted by National Statistical Institutes (NSIs), non-probability data sources are becoming more popular \citep{berkesewicz2017two, beaumont2020probability}. Non-probability surveys, such as opt-in web panels, social media, scanner data, mobile phone data or voluntary register data, are currently being explored for use in the production of official statistics \citep{citro2014multiple,daas2015big}. Since the selection mechanism in these sources is unknown, standard design-based inference methods cannot be directly applied. 

To address this problem, several approaches based on inverse probability weighting (IPW), mass imputation (MI) and doubly robust (DR) estimators have been proposed for two main scenarios: 1) population-level data are available, either in the form of unit-level data (e.g. from a~register covering the whole population) or known population totals/means, and 2) only survey data are available as a~source of information about the target population \citep[cf.][]{elliott_inference_2017}. \citet{wu2022statistical} classified these approaches into three groups that require a~joint randomization framework involving $p$ (probability sampling design) and one of the outcome regression model $\xi$ or propensity score model $q$. In this approach the IPW estimator is under the $qp$ framework, the MI estimator is under the $\xi p$ framework, DR is under the $qp$ or $\xi p$ framework.

Most approaches assume that population data are used to reduce the bias of non-probability sampling by a~proper reweighting to reproduce known population totals/means, by modelling $E(Y|\bX)$ using various techniques or combining both approaches (for instance doubly robust estimators, cf. \citet{chen2020doubly}; multilevel regression and post-stratification (MRP) also called \textit{Mister-P}, cf. \citet{gelman1997poststratification}). Majority of these methods rely on a~limited number of moments of continuous or count data, with some exceptions. For example, non-parametric approaches based on nearest neighbours (NN), such as those discussed by Yang, Kim and Hwang (\citeyear{yang2021integration}) or kernel density estimation (KDE) described by Chen, Yang and Kim (\citeyear{chen_nonparametric_2022}) have also been proposed. In general, the standard approach, such as IPW, seems to be limited, although surveys and register data contain continuous or count data which can be used to account for distribution mismatches. This is because the standard IPW estimators do not allow to match quartiles, deciles or whole distribution between samples.

In this paper, we focus on the $qp$ framework and extend existing IPW estimators to account for distribution differences through quantiles. We use the idea of survey calibration for quantiles described by \citet{harms2006calibration} and propose a~\textit{quantile balancing} IPW (hereinafter QBIPW) estimator that reproduces quantiles and totals jointly for a set of auxiliary variables. Our contribution can be summarised as follows:

\begin{itemize}
    \item we extend IPW estimator for non-probability samples to account not only for totals/means but also for the distribution of auxiliary variables in the form of a~set of quantiles;
    \item we show that existing inference methods hold for the proposed techniques;
    \item we provide proofs for the existence of solutions for the proposed techniques;
    \item we show that adding information on quantiles is equivalent to using piecewise (constant) regression, which can be used to approximate a~non-linear relationship through a~linear model, making IPW more robust to model mis-specification, and consequently, decreasing the bias and improving efficiency of estimation.
\end{itemize}

In this paper we generally follow the notation used by \citet{wu2022statistical}. The paper has the following structure. In Section \ref{sec-basic} we introduce the basic setup: the notation and a~short description of calibration for totals and quantiles separately. In Section \ref{sec-ipw-general} we discuss the IPW estimator for the non-probability samples. In Section \ref{sec-qbipw} we present the proposed estimator. Section \ref{sec-sim-study} contains results of a~simulation study and finally, Section \ref{sec-application} summarises results from a~study in which the proposed approach was applied to combined data from a~job vacancy survey and vacancies from administrative (public employment offices) sources. The paper ends with conclusions and identifies research problems that require further investigation.
 
\section{Basic setup}\label{sec-basic}

\subsection{Notation}

Let $U=\{1,..., N\}$ denote the target population consisting of $N$ labelled units. Each unit $k$ has an associated vector of auxiliary variables $\bx$ and the target variable $y$, with their corresponding values $\bx_k$ and $y_k$, respectively. Let $\{ (y_k, \bx_k), k \in S_A\}$ be a~dataset of a~non-probability sample $S_A$ of size $n_A$ and $\{(\bx_k, d_k^B), k \in S_B\}$ -- a~dataset of a~probability sample $S_B$ of size $n_B$. Only information about auxiliary variables $\bx$ is found in both datasets and $d_k^B = 1/\pi_k^B$ are design-based weights assigned to each unit in the sample $S_B$.  Table \ref{tab-two-sources} summarises the setup, which contains no overlap.

\begin{table}[ht!]
    \centering
    \caption{A setup with two data sources}\label{tab-two-sources}
    \resizebox{\linewidth}{!}{
    \begin{tabular}{llccc}
    \hline
     Sample    & ID & Sample weight $d=\pi^{-1}$ & Covariates $\bx$ & Study variable $y$ \\
    \hline
     Non-probability sample ($S_A)$   & 1 & ? & $\checkmark$ & $\checkmark$ \\
                                      & $\vdots$ & ? & $\vdots$ & $\vdots$ \\
                                      & $n_A$ & ? & $\checkmark$ & $\checkmark$ \\
     Probability sample ($S_B$)      & 1 & $\checkmark$ & $\checkmark$ & ?\\
                                     & $\vdots$ & $\vdots$ & $\vdots$ & ?\\
                                     & $n_B$ & $\checkmark$ & $\checkmark$ & ?\\                                     
    \hline     
    \end{tabular}
}
\end{table}

Let $R_k=I(k \in S_A)$ be the indicator variable showing that unit $k$ is included in the non-probability sample $S_A$, which is defined for all units in the population $U$. Let $\pi_k^A=P(R_k = 1 | \bx_k)$ for $k =1, ..., N$ be their propensity scores. For $\pi_k^A$ we can assume a~parametric model $\pi(\bx_k; \bgamma)$ (e.g. logistic regression) with unknown model parameters $\bgamma$. 

The goal is to estimate a~finite population total $\tau_{y}=\sum_{k\in U}y_{k}$ or the mean $\bar{\tau}_{y}=\tau_{y}/N$ of the variable of interest $y$. In probability surveys (assuming that $y_{k}$ are known), the Horvitz-Thompson is the well-known estimator of a~finite population total, which is expressed as $\hat{\tau}_{y\pi}=\sum_{k\in S_B}{d_{k}^{B}y_{k}}$. This estimator is unbiased for $\tau_{y}$ i.e. $E\left(\hat{\tau}_{y\pi}\right)=\tau_{y}$. However, a~different approach should be applied for non-probability surveys, as $\pi_k^A$ are unknown and have to be estimated (considering also that in our setup $y_{k}$ are unknown in $S_{B}$ and $\hat{\tau}_{y\pi}$ cannot be used). 

In the paper we use the idea of \citet{harms2006calibration} for quantiles which is an extension of the calibration approach proposed by \citet{deville1992calibration} for totals, thus in the next section we briefly discuss these two methods.

\subsection{Calibration estimator for a~total}

Let $\bx$ be a~vector of auxiliary variables (benchmark variables) for which $\tau_{\bx}=\sum_{k\in U}\bx_{k}$ is assumed to be known. The main idea of calibration for probability samples is to look for new calibration weights $w_{k}^{B}$ that are as close as possible to original weights $d_{k}^{B}$ and reproduce known population totals $\tau_{\bx}$ exactly. In other words, in order to find a~new vector of calibration weights ${\bw}=\argmin_{\bv}D\left(\bd,\bv\right)$ we minimise a~distance function $D\left(\bd,\bv\right)=\sum _{k\in S_B}d_{k}^{B}\hspace{2pt} G\hspace{0pt}\left(\frac{v_{k}^{B}}{d_{k}^{B}}\right) \to \min$ subject to calibration equations $\sum_{k\in S_B}v_{k}^{B}\bx_{k} = \sum_{k\in U}\bx_{k}$, where $\bd=\left(d_{1}^{B},\ldots,d_{n_{B}}^{B}\right)^T$, $\bv=\left(v_{1}^{B},\ldots,v_{n_{B}}^{B}\right)^T$ and $G\left(\cdot\right)$ is a~function that must satisfy some regularity conditions. For instance, if $G\left(x\right)=\frac{\left(x-1\right)^{2}}{2}$, then using the method of Lagrange multipliers the final calibration weights $w_{k}^{B}$ can be expressed as $w_{k}^{B}=d_{k}^{B}+d_{k}^{B}\left(\tau_{\bx}-\hat{\tau}_{\bx\pi}\right)^T\left(\sum_{j\in S_B}d_{j}^{B}\bx_{j}\bx_{j}^{ T}\right)^{-1}\bx_{k}$.  The final calibration estimator of a~population total $\tau_{y}$ (assuming that $y_{k}$ would be known in a~sample $S_{B}$) can be expressed as $\hat{\tau}_{y\bx}=\sum_{k\in S_{B}}w_{k}^{}y_{k}$.

\subsection{Calibration estimator for a~quantile} \label{sec2.3}

\citet{harms2006calibration} considered the estimation of quantiles using the calibration approach in a~very similar way to what \citet{deville1992calibration} proposed for a~finite population total $\tau_{y}$. By analogy, in their approach it is not necessary to know values of all auxiliary variables for all units in the population. It is enough to know the corresponding quantiles for the benchmark variables. Below we briefly discuss the~problem of finding calibration weights for probability sample $S_B$.

We want to estimate a~quantile $Q_{y,\alpha}$ of order $\alpha \in \left(0,1\right)$ of the variable of interest $y$, which can be expressed as $Q_{y,\alpha}=\mathrm{inf}\left\{t\left|F_{y}\left(t\right)\geq \alpha \right.\right\}$, where $F_{y}\left(t\right)=N^{-1}\sum_{k\in U}H\left(t-y_{k}\right)$ and the Heaviside function is given by 
\begin{equation}\label{H}
H\left(t-y_{k}\right)=\left\{ \begin{array}{ll}
1, & \ t \geq y_{k},\\  
0, & \ t<y_{k}.\\
\end{array} \right.
\end{equation}

We assume that $\bQ_{\bx,\alpha}$ is a~vector of known population quantiles of order $\alpha$ for a~vector of auxiliary variables $\bx$, where $\alpha \in \left(0,1\right)$. Even though we use the same notation for a~vector ${\bx}$ its elements in calibration approach for totals and quantiles will be different respectively.  In most cases quantiles will be known for continuous auxiliary variables, unlike totals, which will generally be known for categorical variables. It may, however, happen that for a~specific auxiliary variable its population total and the corresponding quantile of order $\alpha$ will be known. In order to find a~new vector of calibration weights ${\bw}$ that reproduces known population quantiles in a~vector $\bQ_{\bx,\alpha}$, an interpolated distribution function estimator of $F_{y}\left(t\right)$ is defined as $
\hat{F}_{y,cal}(t)=\frac{\sum_{k \in S_B} w_{k}^{B} H_{y, S_{B}}\left(t, y_{k}\right)}{\sum_{k \in S_B} w_{k}^{B}} 
$, where the Heaviside function in formula (\ref{H}) is replaced by the modified function $H_{y, S_{B}}\left(t, y_{k}\right)$ given by

\begin{equation}
H_{y, S_{B}}\left(t, y_{k}\right)=\left\{
\begin{array}{ll}
1, & y_{k} \leqslant L_{y, S_{B}}(t), \\ 
\vartheta_{y, S_{B}}\left(t\right), & y_{k}=U_{y, S_{B}}\left(t\right), \\ 
0, & y_{k}>U_{y, S_{B}}\left(t\right),
\end{array}\right. 
\label{eq-hfunction-y}
\end{equation}

\noindent where appropriate parameters are defined as $L_{y, S_{B}}\left(t\right)=\max \left\{\left\{y_{k}, k \in S_{B} \mid y_{k} \leqslant t\right\} \cup\{-\infty\}\right\}$, $U_{y, S_{B}}\left(t\right)=\min \left\{\left\{y_{k}, k \in S_{B} \mid y_{k}>t\right\} \cup\{\infty\}\right\}$ and $\vartheta_{y, S_{B}}\left(t\right)=\frac{t-L_{y, S_{B}}\left(t\right)}{U_{y, S_{B}}\left(t\right)-L_{y, S_{B}}\left(t\right)}$ for $k=1,\ldots,n_{B}$, $t \in \mathbb{R}$. From a~practical point of view a~smooth approximation to the step function, based on the logistic function can be used i.e. $H\left(x\right)\approx\frac{1}{2}+\frac{1}{2}\tanh{kx}=(1+e^{-2kx})^{-1}$, where a~larger value of $k$ corresponds to a~sharper transition at $x = 0$.

A calibration estimator of quantile $Q_{y,\alpha}$ of order $\alpha$ for variable $y$ is defined as $\hat{Q}_{y,cal,\alpha}=\hat{F}_{y,cal}^{-1}(\alpha)$, where a~vector $\bw=\argmin_{\bv}D\left(\bd,\bv\right)$ is a~solution of an optimization problem $D\left(\bd,\bv\right)=\sum _{k\in S_B}d_{k}^{B}\hspace{2pt} G\hspace{0pt}\left(\frac{v_{k}^{B}}{d_{k}^{B}}\right) \to \min$ subject to the calibration constraints $\sum_{k\in S_B}v_{k}^{B}=N$ and  $\hat{\bQ}_{\bx,cal,\alpha}=\bQ_{\bx,\alpha}$ or equivalently $\hat{F}_{x_{j},cal}\left(Q_{x_{j},\alpha}\right)=\alpha$ for all auxiliary variables $x_{j}$ from a~vector $\bx$.

As in the previous case, if $G\left(x\right)=\frac{\left(x-1\right)^{2}}{2}$, then, using the method of Lagrange multipliers, the final weights $w_{k}^{B}$ can be expressed as $w_{k}^{B}=d_{k}^{B}+d_{k}^{B}\left(\mathbf{T_{a}}-\sum_{k\in S_B}{d_{k}^{B}\ba_{k}}\right)^{T}\left(\sum_{j\in S_B}{d_{j}^{B}}\ba_{j}\ba_{j}^{T}\right)^{-1}\ba_{k}$, where $\mathbf{T_{a}}=\left(N,\alpha,\ldots,\alpha\right)^{T}$ and the elements of $\ba_{k}$ (the first element is 1) are given by

\begin{equation}
a_{kj}=\left\{\begin{array}{lll} 
N^{-1},& \quad x_{kj}\leq L_{x_{j},S_{B}}\left(Q_{x_{j},\alpha}\right),\\
N^{-1}\vartheta_{x_{j},S_{B}}\left(Q_{x_{j},\alpha}\right), & \quad x_{kj}=U_{x_{j},S_{B}}\left(Q_{x_{j},\alpha}\right),\\
0,& \quad x_{kj}> U_{x_{j},S_{B}}\left(Q_{x_{j},\alpha}\right).\\
\end{array} \right.
\label{cal-var-a}
\end{equation}

In the method described above it is assumed that a~known population quantile is reproduced for a~set of auxiliary variables, i.e. that the process of calibration is based on a~particular quantile (of order~$\alpha$). For instance, it could be the median $\alpha=0.5$. It is a~straightforward process to find calibration weights that accurately reflect population quantiles (for instance, quartiles) for a~selected set of auxiliary variables.

\section{Inverse probability weighting estimator}\label{sec-ipw-general}

\subsection{Assumptions}\label{sec-ipw-standard-assump}

Throughout the paper we follow standard assumptions for inference based on non-probability samples under the $qp$ framework, i.e:

\begin{itemize}
    \item[(A1)] conditional independence of $R_k$ and $y_k$ given $\bx_k$; 
    \item[(A2)] all units in the target population have non-zero propensity scores $\pi_k^A > 0$; 
    \item[(A3)] $R_1,...,R_N$ are independent given auxiliary variables $(\bx_1, ..., \bx_N)$. 
\end{itemize}

To estimate the variance of the IPW estimator we follow the same regularity conditions as stated in the Supplementary Materials in the Chen, Li and Wu (\citeyear{chen2020doubly}) paper and Section 3.2 of \citet{tsiatis2006semiparametric}. For simplicity we will use $\bx_k$ (for which totals are known or estimated) and $\ba_k$ (for which specific $\alpha$-quantiles of selected variables in $\bx_k$ are known or estimated) in the definitions of the IPW estimators.

\subsection{The IPW estimator}

\citet{chen2020doubly} discussed inverse probability weighting (IPW) and its asymptotic properties. Let us assume that propensity scores can be modelled parametrically as $\pi_k^A = P(R_k=1 | \bx_k) = \pi(\bx_k; \bgamma_0)$, where $\bgamma_0$ is the true value of unknown model parameters. The maximum likelihood estimator of $\pi_k^A$ is computed as $\hat{\pi}_k^A = \pi(\bx_k; \hat{\bgamma})$, where $\hat{\bgamma}$ maximises the log-likelihood function under full information:

$$
\ell(\bgamma)=\log \left\{\prod_{k=1}^{N}\left(\pi_{k}^{A}\right)^{R_{k}}\left(1-\pi_{k}^{A}\right)^{1-R_{k}}\right\}=
\sum_{k \in S_{A}} \log \left(\frac{\pi_{k}^{A}}{1-\pi_{k}^{A}}\right)+\sum_{k \in U} \log \left(1-\pi_{k}^{A}\right).
$$

However, in practice, reference auxiliary variables $\bx$ can be supplied by the probability sample $S_B$, so $\ell(\bgamma)$ is replaced with pseudo-likelihood (PL) function:

$$
\ell^{*}(\bgamma)=\sum_{k \in S_{A}} \log \left(\frac{\pi_{k}^{A}}{1-\pi_{k}^{A}}\right)+\sum_{k \in S_{B}} d_{k}^{B} \log \left(1-\pi_{k}^{A}\right).
$$

The maximum PL estimator $\hat{\bgamma}$ can be obtained by solving pseudo-score equations given by $\bU(\bgamma) = \partial \ell^{*}(\bgamma)/\partial \bgamma = \bZero$. If logistic regression is assumed for $\pi_k^A$, then $\bU(\bgamma)$ is given by

\begin{equation}
\bU(\bgamma)=\sum_{k \in S_{A}} \bx_{k}-\sum_{k \in S_{B}} d_{k}^{B} \pi\left(\bx_{k}; \bgamma\right) \bx_{k}.
\label{eq-ipw-u}    
\end{equation}

Note that weights based on $\pi\left(\bx_{k}; \hat{\bgamma}\right)$ where $\hat{\bgamma}$ are estimated based on \eqref{eq-ipw-u} do not reproduce the population size nor totals for variables in a~vector $\bx$. 

Alternatively, pseudo-score equations $\bU(\bgamma)=\bZero$ can be replaced by a~system of general estimating equations \citep{kim_theory_2012, wu2022statistical}. Let $\bh(\bx_k; \bgamma)$ be a~user-specified vector of functions with the same dimensions as $\bgamma$. The general estimating equations are given by 

\begin{equation}
\bG(\bgamma)=\sum_{k \in S_{A}} \bh\left(\bx_{k}; \bgamma\right)-\sum_{k \in S_{B}} d_{k}^{B} \pi\left(\bx_{k}; \bgamma\right) \bh\left(\bx_{k}; \bgamma\right), 
\label{eq-ipw-gee}
\end{equation}

\noindent and when we let $\bh(\bx_k; \bgamma) = \bx_k/\pi(\bx_k; \bgamma)$, then \eqref{eq-ipw-gee} reduces to calibration equations:

\begin{equation}
\bG(\bgamma)=\sum_{k \in S_{A}} \frac{\bx_{k}}{\pi\left(\bx_{k}; \bgamma\right)}-\sum_{k \in S_{B}} d_{k}^{B} \bx_{k}.
\label{eq-ipw-gee-totals}
\end{equation}

In this setting unit-level data are not required to get  $\sum_{k \in S_{B}} d_{k}^{B} \bx_{k}$. These sums can be replaced with known or estimated population totals from the reference probability survey. \citet{kim_theory_2012} showed that estimator \eqref{eq-ipw-gee-totals} leads to optimal estimation when a~linear regression model holds for $y$ and $\bx$.

After estimating $\bgamma$ we obtain propensity score $\hat{\pi}_k^A$ and two versions of the IPW estimator: 

\begin{equation} 
\hat{\bar{\tau}}_{\mathrm{IPW} 1}=\frac{1}{N} \sum_{k \in S_{A}} \frac{y_{k}}{\hat{\pi}_{k}^{A}} \quad \text { and } \quad \hat{\bar{\tau}}_{\mathrm{IPW} 2}=\frac{1}{\hat{N}_{A}} \sum_{k \in S_{A}} \frac{y_{k}}{\hat{\pi}_{k}^{A}},  
\label{eq-ipw-estimators}
\end{equation}

\noindent where $\hat{N}_A = \sum_{k \in S_A} \hat{\pi}_k^A$.

The standard IPW estimator takes only limited information into account and does not preserve the distribution of auxiliary variables. To overcome this limitation we use the idea of the calibration approach for quantiles described in Section \ref{sec2.3} and modify the IPW estimator to balance distributions via specific quantiles (e.g. quartiles, deciles, etc.).

\section{Quantile balancing inverse probability weighting}\label{sec-qbipw}

\subsection{Proposed approach}

In our proposal we modify the propensity scores $\pi_k^A$ to take into account $\ba_k$ in such way that the function will be given by $\pi_k^A = \pi(\bx_k, \ba_k; \bEta)$. The proposed approach can be justified by the fact that the parametric form $\pi\left(\bx_k, \ba_k; \bEta\right)$ is an over-saturated representation that includes $\pi\left(\bx_k; \bgamma\right)$ as a~special case. In addition, note that we distinguish $\bgamma$ from $\bEta$ as the latter contains additional parameters referring to $\ba_k$. Inclusion of $\ba_k$ allows to correct the discrepancies between quantiles of auxiliary variables of $S_A$ with the reference to $S_B$. 

The proposed approach is similar to those used in the literature on survey sampling or causal inference, which offer arguments for the inclusion of higher moments for $\bx$ auxiliary variables (Ai, Linton \& Zhang, \citeyear{ai_simple_2020}; Imai \& Ratkovic, \citeyear{imai_covariate_2014}). On the other hand, it is a~simpler version of the propensity score that balances distributions in comparison to methods using kernel density \citep{hazlett_kernel_2020} or numerical integration (Sant'Anna et al., \citeyear{sant2022covariate}).

Our approach is not limited to a~specific estimation method of the parameters of the propensity score $\pi_k^A$. Thus, we start with the use of the full or pseudo-likelihood function with the proposed modification and then solve it using (\ref{eq-ipw-u}), which takes the following form if both $\bx_k$ and $\ba_k$ are used

\begin{equation}
\bU(\bgamma) = 
\begin{pmatrix}
    \sum_{k \in S_{A}} \bx_{k}-\sum_{k \in S_{B}} d_{k}^{B} \pi\left(\bx_{k}, \ba_k; \bEta\right) \bx_{k} \\
    \sum_{k \in S_{A}} \ba_{k}-\sum_{k \in S_{B}} d_{k}^{B} \pi\left(\bx_{k}, \ba_{k};\bEta\right) \ba_{k}
\end{pmatrix}.
\label{eq-mle-ipw}
\end{equation}

Note that this procedure does not allow to reproduce quantiles for the selected $\bx$ variables but it reduces the discrepancies between distributions of $S_A$ and $S_B$. In the simulation study we show that this is particularly significant for the non-linear selection models. On the other hand, including more variables may lead to instability of the estimation procedure and large weights.

To overcome the limitation of maximum likelihood estimation (MLE) we can use the generalised estimating equations in \eqref{eq-ipw-gee} which allow to reproduce totals and quantiles from the probability sample $S_B$. In this situation $\bG(\bEta)$ changes to

$$
\begin{aligned}
   \bG(\bEta)= &  \sum_{k \in S_{A}} \bh\left(\bx_{k}, \ba_k; \bEta\right)-\sum_{k \in S_{B}} d_{k}^{B} \pi\left(\bx_{k}, \ba_k; \bEta\right) \bh\left(\bx_{k}, \ba_k; \bEta\right).
\end{aligned}
$$

If we assume logistic regression for $\pi\left(\bx_k, \ba_k;  \bEta\right)$ and $\bh(\bx_k, \ba_k; \bEta) = \left(\begin{smallmatrix}\bx_{k}\\\ba_{k}\end{smallmatrix}\right)/\pi(\bx_k, \ba_k;\bEta)$ then $\bG(\bEta)$ reduces to 

\begin{equation}
\bG(\bgamma)=
\begin{pmatrix}
\sum_{k \in S_{A}} \frac{\bx_{k}}{\pi\left(\bx_{k}, \ba_k; \bEta\right)}-\sum_{k \in S_{B}} d_{k}^{B} \bx_{k} \\
\sum_{k \in S_{A}} \frac{\ba_{k}}{\pi\left(\bx_{k},\ba_k; \bEta\right)}-\sum_{k \in S_{B}} d_{k}^{B} \ba_{k}
\end{pmatrix}.
\label{eq-gee-ipw}
\end{equation} 

This allows to estimate the $\bEta$ parameters so that the population size $N$, known population totals (or estimated population totals) for some auxiliary variables in a vector $\bx$ or/and known population quantiles for other variables from this vector are all being reproduced. In the Section \ref{appen-existance} in the Appendix we state conditions which guarantee existence and uniqueness of solutions for $\bU(\bEta)=\boldsymbol{0}$ and $\bG(\bEta)=\boldsymbol{0}$.

\begin{remark}
\citet[p. 41]{harms2006calibration} note that if the relationship between $y$ and a~scalar auxiliary variable $x$ is exactly linear then calibration to an $\alpha$-quantile of the auxiliary variable the yields exact population $\alpha$-quantile of the target variable $y$. 
\end{remark}

\begin{remark}
The use of $\ba_k$ in \eqref{eq-gee-ipw} is related to the use of piecewise (constant) regression where variables $\bx_k$ are split into breaks (e.g. by quartiles or deciles) and then $\ba_k$ are included in linear regression as given below

$$
\hat{y}_k = 
\underbrace{\bx_k^T\hat{\bbeta}}_{\text{linear part}} + 
\underbrace{\ba_k^T\hat{\bbeta}^*.}_{\text{piecewise part}}
$$

Thus, we use the piecewise method to approximate the non-linear relationship between $y$ and $\bx$. Consider the following example: generate $n=1,000$ observations of $X \sim \Uni(0,80)$, $Y_1 \sim N(1300-(X-40)^2, 300)$ and $p=\text{P}(Y_2=1)=\logit(-3+(X-1.5)^2 + N(0, 0.5))$ to yield $(y_k, p_k, x_k)$ triplet for $k=1,...,n$. Then, let $x_k$ be an auxiliary variable in the following settings: 1) $x_k$ is used as it is and 2) $x_{k}$ is used along with $a_k$ specified for quartiles, and 3) $x_{k}$ is used along with $a_k$ specified for deciles. Figure \ref{fig-example1} presents how the prediction of $Y_1$ (first row) and $\text{P}(Y_2=1)$ (second row) changes when $x_k$ or $x_k$ and $a_k$ are used. For $Y_1$ we used linear regression and for $Y_2$ probabilities we modelled using logistic regression. As expected including quantiles through $\ba_k$ improves prediction of both models.

\begin{figure}[ht!]
    \centering
    \includegraphics[width=0.9\textwidth]{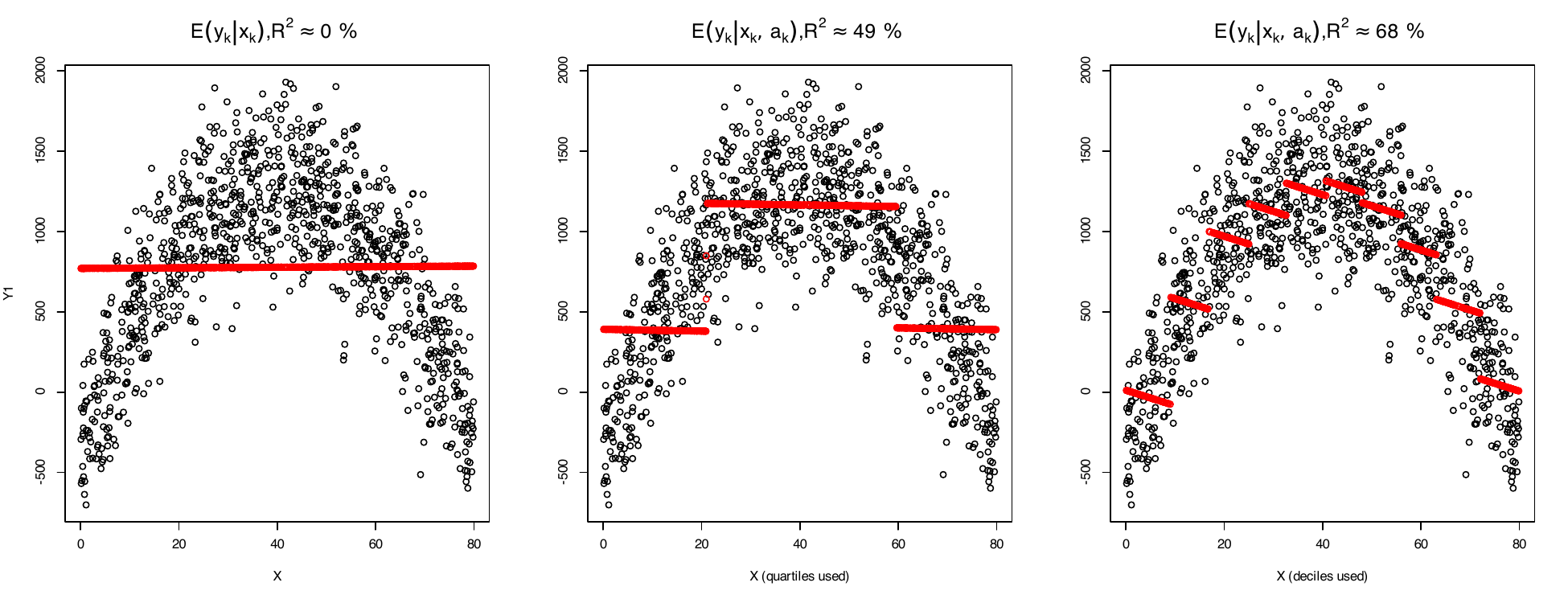}
    \includegraphics[width=0.9\textwidth]{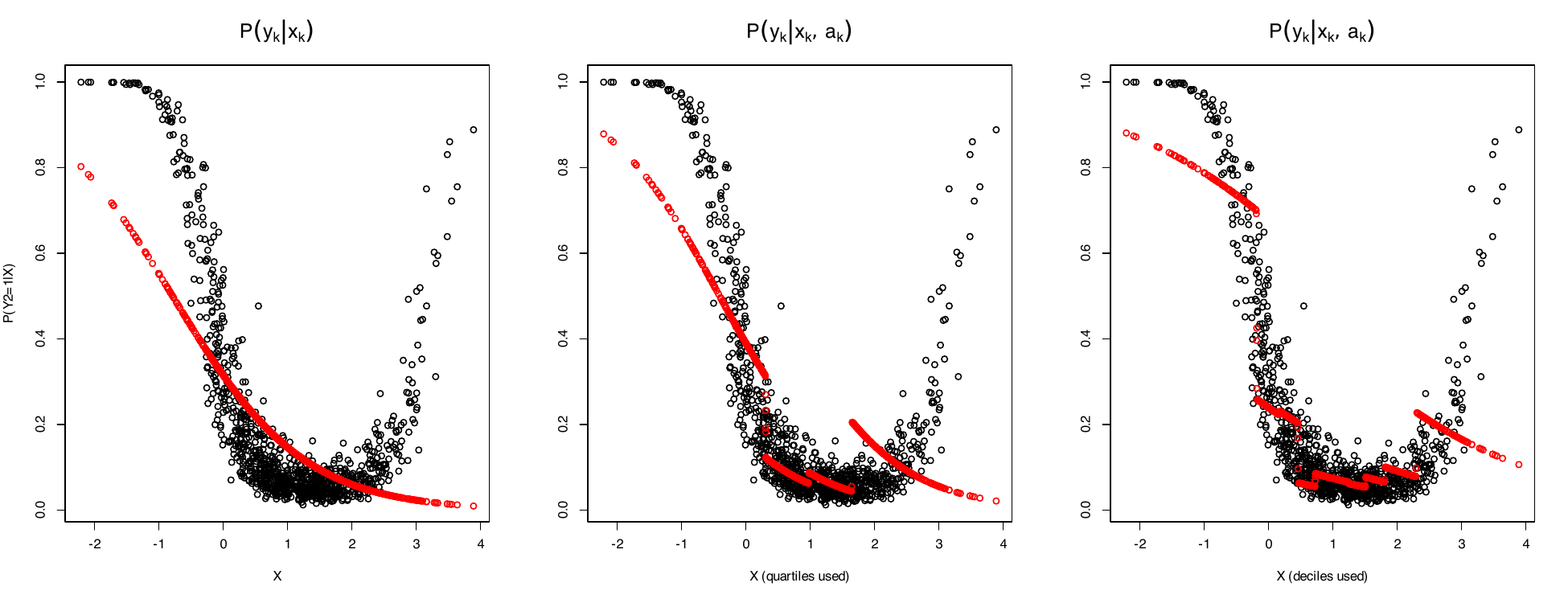}
    \caption{Prediction of $Y_1$ (top row) and $\text{P}(Y_2=1)$ (bottom row) based on $x_k$ or $x_k$ and $a_k$ (quartiles or deciles) from the example in Remark 2. Black circles denote observed $y_k$, red circles denote predictions $\hat{m}_k$ based on the linear model and $\hat{p}_k$ based on logistic regression} 
    \label{fig-example1}
\end{figure}
    
\end{remark}

\begin{remark}
The QBIPW estimator obtained by solving \eqref{eq-gee-ipw} is actually doubly robust i.e. it is asymptotically unbiased if either the inverse probability model or outcome model is correct. Usage of quantiles through $\ba_k$ allows to approximate non-linear relationships in both models. In the simulation study we will show that the proposed approach significantly reduces bias when the two models are non-linear.
\end{remark}

After estimation of the $\bEta$ parameters we can estimate the population mean using either $\hat{\bar{\tau}}_{IPW1}$ or $\hat{\bar{\tau}}_{IPW2}$ given in equation \eqref{eq-ipw-estimators} which we denote as $\hat{\bar{\tau}}_{QBIPW1}$ or $\hat{\bar{\tau}}_{QBIPW2}$ respectively. To summarize our proposed approach:

\begin{itemize}
    \item estimated propensity scores will reproduce either known or estimated moments as well as specified $\alpha$-quantiles (e.g. median, quartiles or deciles, etc.). We use more information regarding count or continuous variables if they are available in both datasets;
    \item inclusion of $\alpha$-quantiles in the $\pi_k^A$ approximates the relationship between inclusion $R$ and $\bx$. Thus, the inclusion of $\alpha$-quantiles makes estimates more robust to model mis-specification for the propensity and outcome model when calibrated IPW is used.
\end{itemize}

Inference based on the IPW estimator does not change as we add new variables to the propensity score model. We do not assume that quantiles are based on, let's say, instrumental variables or are collected through paradata (Park, Kim \& Kim, \citeyear{park_note_2019}) but treat them as if we are adding higher moments. Estimation of the the variance of the QBIPW estimators is provided in the next section.

\subsection{Variance estimation}

In order to estimate the variance of the QBIPW estimators 
one can modify the equation $\bPhi_n(\bEta)=\bZero$ \citep[][eq. 6.1]{wu2022statistical} by imposing constraints on quantiles. In the case when both $\bx_k$ and $\ba_k$ are used and $\bh(\bx_k,\ba_k; \bEta)$ is defined as previously, the equation is given as: 

\begin{equation}
\bPhi_n(\bEta)=N^{-1}
\begin{pmatrix}
\sum_{k \in U} R_k\frac{y_k-\bar{\tau}_y}{\pi(\bx_k, \ba_k; \bEta)} \\
\sum_{k \in U} R_k \frac{\bx_k}{\pi(\bx_k, \ba_k; \bEta)} - \sum_{k \in S_B} d_k^B \bx_k \\
\sum_{k \in U} R_k \frac{\ba_k}{\pi(\bx_k, \ba_k; \bEta)} - \sum_{k \in S_B} d_k^B \ba_k
\end{pmatrix}.
\end{equation}

The asymptotic variance for the QBIPW estimator can be derived from the standard sandwich form given by:

$$
\operatorname{AV}(\hat{\bEta})=\left[E\left\{\boldsymbol{\phi}_n\left(\bEta_0\right)\right\}\right]^{-1} \operatorname{Var}\left\{\boldsymbol{\Phi}_n\left(\bEta_0\right)\right\}\left[E\left\{\boldsymbol{\phi}_n\left(\bEta_0\right)\right\}^{\prime}\right]^{-1},
$$

\noindent where $\bEta_0$ is the true value of unknown model parameters and $\boldsymbol{\phi}_n(\bEta)=\partial \boldsymbol{\Phi}_n(\bEta) / \partial \bEta$ depends on the form of $\pi_k^A$ and $\bh$. \citet{chen2020doubly} provided the analytical form of the variance estimator when logistic regression is used for $\pi_k$. 

Our approach, based on $\ba_k$ or $\bx_k$, may lead to higher variance as the number of variables can be significantly larger than $\bx_k$ alone as adding new variables may inflate the variance of $\pi_k^A(1-\pi_k^A)$.

\section{Simulation study}\label{sec-sim-study}

Our simulation study to compare the proposed estimators follows the procedure described in \citet{kim2019sampling} and Yang, Kim and Hwang (\citeyear{yang2021integration}). We generate a~finite population $\mathcal{F}_N = \{\bx_k = (x_{k1}, x_{k2}), \by_k = (y_{k1}, y_{k2}): k = 1,\ldots,N\}$ with size $N=100,000$, where $y_{k1}$ is the continuous outcome, $y_{k2}$ is the binary outcome, $x_{k1} \sim N(1,1)$ and $x_{k2} \sim \Exp(1)$. 

The finite population target variables ($Y_1$ and $Y_2$) and the big sample $S_A$ were generated using the outcome variables (denoted as outcome model; OM) and the inclusion probability (denoted as probability model; PM) presented in Table~\ref{tab-om-pm-defs}, where $\alpha_k \sim N(0,1)$, $\epsilon_k\sim N(0,1)$ and $x_{k1}, x_{k2}, \alpha_k$ and $\epsilon_k$ are mutually independent. The variable $\alpha_k$ induces dependence of $y_{k1}$ and $y_{k2}$ even after adjusting for $x_{k1}$ and $x_{k2}$.

\begin{table}[ht!]
    \centering
    \caption{Outcome models and the probability of inclusion in sample $S_A$ used in the simulation study}
       \label{tab-om-pm-defs}
       \resizebox{\linewidth}{!}{
    \begin{tabular}{lll}
    \hline
    Type & Form & Formulae \\
    \hline
    Continuous & linear (OM1) &  $y_{k1} = 1 + x_{k1} + x_{k2} + \alpha_k + \epsilon_k $\\
             & non-linear (OM2) & $y_{k2}= 0.5(x_{k1}-1.5)^2 + x_{k2}^2 + \alpha_k + \epsilon_k$ \\ 
    Binary  & linear (OM3) &  $P(y_{k1} = 1 | x_{k1}, x_{k2}; \alpha_k) = \logit(1 + x_{k1} + x_{k2} + \alpha_k)$ \\
            & non-linear (OM4) &  $P(y_{k2}=1|x_{k1}, x_{k2}; \alpha_k) = \logit\left\{0.5(x_{k1}-1.5)^2 + x_{k2}^2 + \alpha_k\right\}$\\
    Selection ($S_A$) & linear (PM1) & $\logit(p_k)=x_{k2}$\\
      & non-linear (PM2) & $\logit(p_k)= -3 + (x_{k1}-1.5)^2 + (x_{k2}-2)^2$\\
    \hline       
    \end{tabular}
    }
\end{table}

From this population, we select a~big sample $S_A$ of size approximately $70,000$ (for PM1) and $50,000$ (for PM2), depending on the selection mechanism, assuming the inclusion indicator $\delta_{Bk} \sim \text{Bernoulli}(p_k)$ with $p_k$ denoting the inclusion probability for unit $k$. Then we select a~simple random sample $S_B$ of size $n=1,000$ from $\mathcal{F}_N$. The goal is to estimate the population mean $\bar{\tau}_j = N^{-1}\sum_{k=1}^Ny_{kj}$, $j=1,2$.

In the results, we report Monte Carlo bias (B), standard error (SE) and root mean square error (RMSE)  based on $R=500$ simulations for each $y$ variable: $\text{B}  = \bar{\hat{\tau}} - \bar{\tau}$, $ \text{SE}= \sqrt{\frac{\sum_{r=1}^R \left(\hat{\tau}^{(r)} - \bar{\hat{\tau}}\right)^2}{R-1}}$ and $\text{RMSE} = \sqrt{ \text{B}^2 + \text{SE}^2}$, where $\bar{\hat{\tau}} = \sum_{r=1}^{R}\hat{\tau}^{(r)} / R$ and $\hat{\tau}^{\left(r\right)}$ is an estimate of the mean in the $r$-th replication. The simulation was conducted in R \citep{r-cran} using  the \texttt{nonprobsvy} \citep{nonprobsvy} and the \texttt{jointCalib} \citep{jointcalib} packages. To solve calibration constraints in the QBIPW estimator we used Newton's method with a~trust region global strategy \textit{double deglog} proposed by \citet{dennis1996numerical} and implemented in the \texttt{nleqslv} \citep{nleqslv} package.  
 
The simulation results are presented in Tables \ref{tab-results} and  \ref{coverage}. The latter shows the empirical coverage (CR) of 95\% confidence intervals (CI) based on analytical variance estimators for the IPW estimators. The CR is defined as $\frac{1}{R} \sum_{r=1}^R I\left(\bar{\tau} \in \mathrm{CI}^{(r)}\right) \times 100$, where $\mathrm{CI}^{(r)}$ is the CI calculated in the $r$-th iteration, and we expect it to be close to the nominal 95\% level. Table \ref{tab-quality} in the appendix also provides information on the quality of the sums and quantiles reproduced.

The following estimators were considered with a~set of calibration equations including totals or quantiles for auxiliary variables estimated from sample $S_B$:

\begin{itemize}
    \item Na\"ive calculated from sample $S_A$ only,
    \item Mass imputation nearest neighbour with 5 neighbours using $x_1, x_2$ only (NN),
    \item Mass imputation based on linear regression using $x_1, x_2$ only (GLM),
    \item Doubly robust estimator using $x_1, x_2$ only with two variants of IPW: MLE (DR MLE; where IPW MLE is based on \eqref{eq-ipw-u}) and GEE (calibrated; DR GEE; where IPW GEE is based on \eqref{eq-ipw-gee-totals}),
    \item IPW MLE and IPW GEE with the following variants: estimated totals for $x_1, x_2$ only,
    \item QBIPW MLE and QBIPW GEE based on (\ref{eq-mle-ipw}) and (\ref{eq-gee-ipw}) respectively with the following variants: 
    \begin{itemize}
        \item estimated quartiles and totals for $x_1, x_2$ (QBIPW1),
        \item estimated deciles and totals for $x_1, x_2$ (QBIPW2).
    \end{itemize}
\end{itemize}

Table \ref{tab-results} contains the results for four scenarios, in which combinations of OM and PM for continuous and binary outcome models are considered, as defined in Table \ref{tab-om-pm-defs}. In Scenario I, for continuous OM, the majority of estimators demonstrate satisfactory performance, exhibiting minimal bias. The proposed estimators, which employ quantiles and totals, demonstrate comparable root mean square error (RMSE) to that of the nearest neighbour (NN), generalised linear model (GLM) or double robust (DR) estimators. As anticipated, the QBIPW with GEE is distinguished by a~lower standard error compared to MLE. For binary outcomes, the performance of the QBIPW estimators, particularly the GEE-based ones, is even more pronounced. The RMSE is comparable to that of GLM and DR estimators. Furthermore, the proposed QBIPW under GEE demonstrates an improvement in RMSE compared to the IPW GEE estimator.

\begin{table}[ht!]
\centering
\scriptsize
\caption{Results for continuous and binary Y (B, SE and RMSE are multiplied by 100)}
\label{tab-results}
\begin{tabular}{lrrrrrrrrrrrr}
  \hline
  & \multicolumn{3}{c}{Scenario I} & \multicolumn{3}{c}{Scenario II}  & 
  \multicolumn{3}{c}{Scenario III}  & \multicolumn{3}{c}{Scenario IV}  \\ 
 OM & \multicolumn{3}{c}{linear (1)} & \multicolumn{3}{c}{linear (1)} & 
      \multicolumn{3}{c}{non-linear (2)} & \multicolumn{3}{c}{non-linear (2)} \\ 
 PM & \multicolumn{3}{c}{linear (1)} & \multicolumn{3}{c}{non-linear (2)} & 
     \multicolumn{3}{c}{linear (1)} & \multicolumn{3}{c}{non-linear (2)} \\ 
 & B & SE & RMSE & B & SE & RMSE & B & SE & RMSE & B & SE & RMSE \\ 
  \hline
  \hline
  \multicolumn{13}{c}{\textbf{Continuous Y (OM1 and OM2)}}\\
  \hline
  \hline
  Naive & 18.71 & 0.40 & 18.70 & -41.84 & 0.50 & 41.80 & 60.67 & 0.60 & 60.70 & 31.11 & 0.80 & 31.10 \\ 
  \hline
  NN & 0.20 & 6.10 & 6.10 & 0.66 & 6.30 & 6.40 & 0.43 & 15.20 & 15.20 & 0.90 & 14.90 & 14.90 \\ 
  GLM & 0.43 & 4.50 & 4.60 & -0.19 & 4.50 & 4.50 & -19.92 & 13.60 & 24.10 & 110.72 & 15.00 & 111.70 \\ 
  DR (MLE) & 0.32 & 4.60 & 4.60 & 0.01 & 4.60 & 4.60 & 1.72 & 5.80 & 6.00 & 227.42 & 77.00 & 240.10 \\ 
  DR (GEE) & 0.33 & 4.60 & 4.60 & -0.19 & 4.50 & 4.50 & 0.96 & 7.10 & 7.10 & 123.79 & 18.10 & 125.10 \\ 
  \hline
  \multicolumn{13}{c}{Inverse probability weighting (MLE)}\\
  \hline
  IPW & -0.01 & 5.70 & 5.70 & 69.73 & 27.50 & 75.00 & 0.42 & 9.30 & 9.40 & 467.97 & 175.10 & 499.60 \\  
  QBIPW1 & -0.02 & 5.70 & 5.70 & 6.58 & 11.70 & 13.50 & 0.44 & 9.60 & 9.60 & 29.93 & 22.20 & 37.30 \\ 
  QBIPW2 & -0.50 & 5.80 & 5.80 & 3.12 & 11.90 & 12.30 & 0.73 & 12.70 & 12.70 & 4.04 & 21.50 & 21.80 \\ 
 \hline
  \multicolumn{13}{c}{Inverse probability weighting (calibrated; GEE)}\\
  \hline
  IPW & 0.33 & 4.60 & 4.60 & -0.19 & 4.50 & 4.50 & 0.96 & 7.10 & 7.10 & 123.79 & 18.10 & 125.10 \\ 
  QBIPW1 & 0.33 & 4.50 & 4.50 & 0.38 & 4.60 & 4.60 & 2.61 & 9.70 & 10.00 & 27.84 & 12.50 & 30.50 \\ 
  QBIPW2 & 0.97 & 4.20 & 4.30 & 0.11 & 4.50 & 4.50 & 4.78 & 10.70 & 11.70 & 8.24 & 12.50 & 14.90 \\ 
  \hline
   \hline
  \multicolumn{13}{c}{\textbf{Binary Y (OM3 and OM4)}}\\
    \hline
    \hline
Naive & 1.04 & 0.07 & 1.04 & -4.19 & 0.07 & 4.19 & 2.85 & 0.09 & 2.85 & -1.51 & 0.10 & 1.52 \\ 
  \hline
  NN & 0.11 & 0.99 & 0.99 & 0.10 & 1.00 & 1.00 & 0.03 & 1.43 & 1.43 & 0.02 & 1.37 & 1.37 \\ 
  GLM & 0.02 & 0.34 & 0.34 & 0.02 & 0.33 & 0.33 & -0.19 & 0.50 & 0.54 & 2.40 & 0.45 & 2.44 \\ 
  DR (MLE) & 0.03 & 0.34 & 0.34 & 0.05 & 0.34 & 0.34 & 0.06 & 0.41 & 0.41 & 3.31 & 0.43 & 3.33 \\ 
  DR (GEE) & 0.03 & 0.34 & 0.34 & 0.03 & 0.33 & 0.33 & 0.04 & 0.42 & 0.42 & 2.78 & 0.42 & 2.82 \\ 
  \hline
  \multicolumn{13}{c}{Inverse probability weighting (MLE)}\\
  \hline
  IPW & -0.02 & 0.47 & 0.47 & 0.37 & 0.70 & 0.79 & -0.02 & 0.77 & 0.77 & 2.32 & 1.31 & 2.66 \\ 
  QBIPW1 & -0.02 & 0.47 & 0.47 & -0.54 & 0.60 & 0.81 & -0.02 & 0.75 & 0.75 & -0.06 & 0.75 & 0.76 \\ 
  QBIPW2 & -0.08 & 0.50 & 0.51 & 0.07 & 0.81 & 0.81 & -0.14 & 0.79 & 0.80 & -0.09 & 1.19 & 1.19 \\ 
 \hline
  \multicolumn{13}{c}{Inverse probability weighting (calibrated; GEE)}\\
  \hline
  IPW  & 0.01 & 0.39 & 0.39 & -1.86 & 0.23 & 1.87 & 0.03 & 0.61 & 0.61 & -0.88 & 0.28 & 0.92 \\ 
  QBIPW1 & -0.01 & 0.37 & 0.37 & -0.54 & 0.25 & 0.60 & -0.02 & 0.58 & 0.58 & 0.09 & 0.52 & 0.52 \\ 
  QBIPW2 & 0.03 & 0.35 & 0.35 & -0.11 & 0.27 & 0.29 & 0.05 & 0.52 & 0.53 & -0.05 & 0.57 & 0.57 \\ 
  \hline
\end{tabular}
\end{table}

In Scenario II, where the OM is linear and the PM is non-linear, it is evident that there are varying performance patterns across estimators. In the case of continuous outcomes, the NN, GLM, and DR estimators demonstrate consistent and reliable performance, exhibiting minimal bias and low RMSE. Notwithstanding, the IPW MLE estimator exhibits a~pronounced increase in both bias and RMSE, underscoring its sensitivity to non-linear probability models. The proposed QBIPW estimators demonstrate superior performance compared to the standard IPW, with QBIPW2 (GEE) exhibiting the lowest bias and a~reasonable SE among IPW estimators. In the case of binary outcomes, the impact of the non-linear PM is less pronounced. The majority of estimators demonstrate low bias, with the QBIPW estimators under GEE exhibiting particularly favourable performance. It is noteworthy that QBIPW2 (GEE) achieves the lowest RMSE (0.29) among all IPW estimators, outperforming even the standard IPW (GEE) estimator.

The third scenario introduces a~non-linear OM with a~linear PM, which presents particular challenges for parametric methods. In the case of continuous outcomes, the GLM estimator demonstrates an elevated level of bias and root mean square error, which is indicative of its inherent limitation in accurately capturing non-linear relationships. The NN estimator demonstrates a~tendency to maintain low bias, although this is accompanied by an increase in SE. The performance of DR methods varies. The DR (MLE) method demonstrates higher bias (1.72) in comparison to the DR (GEE) method (0.96). Among IPW estimators, those belonging to the QBIPW category demonstrate improved performance, with QBIPW2 (GEE) achieving the lowest RMSE within this category. In the case of binary outcomes, the impact of the non-linear OM is less pronounced. The majority of estimators demonstrate low bias, with QBIPW estimators under GEE once again exhibiting robust performance. QBIPW2 (GEE) achieves the lowest RMSE (0.53) among IPW estimators, exhibiting a~performance level comparable to that of GLM and DR estimators.

In Scenario IV, which combines non-linear OM and PM, all estimators are confronted with the most challenging conditions. In the case of continuous outcomes, the majority of estimators are unable to perform satisfactorily. The GLM and DR estimators demonstrate a~considerable degree of bias and error. The NN estimator demonstrates a~tendency to maintain low bias, although it exhibits a~high standard error. Among IPW estimators, those based on maximum likelihood estimation (MLE) exhibit considerable bias and root-mean-square error. The proposed QBIPW estimators under MLE and GEE demonstrate markedly improved performance, with QBIPW2 achieving the lowest bias and RMSE among all estimators. In the case of binary outcomes, although the overall performance is superior to that observed for continuous outcomes, challenges remain. The GLM and DR estimators exhibited an increased bias, with values ranging from 2.40 to 3.31. The proposed QBIPW estimators under GEE once again demonstrate superior performance compared to other IPW estimators. QBIPW2 achieves the lowest RMSE (0.57) in this category, comparable to the NN estimator (RMSE: 1.37). This scenario highlights the effectiveness of the proposed QBIPW estimators, particularly under GEE, in addressing complex non-linear relationships in both outcome and probability models.

Table \ref{coverage} presents the empirical coverage rates (CR) for the proposed estimators. In Scenario I, the proposed estimators exhibited a~CR that was close to the nominal 95\% level. In Scenario II, this phenomenon is particularly evident in the IPW GEE, where a~non-linear propensity score is employed and the assumed model is mis-specified. Nevertheless, the proposed QBIPW2 (with deciles and totals) exhibits CRs that are nearly at the nominal level for both continuous and binary outcomes. Notably, the proposed approach significantly enhances the performance of the standard IPW when MLE is utilized as the estimation technique. The CI has the shortest width for the DR and IPW GEE estimators, which are one third (for continuous) or half (binary) the width of the NN estimator.

\begin{table}[ht]
\centering
\caption{Empirical coverage rate and average length of confidence interval based on analytical variance estimator}\label{coverage}
\scriptsize
\begin{tabular}{llrrrrrrrr}
  \hline
 & & \multicolumn{2}{c}{Scenario I} & \multicolumn{2}{c}{Scenario II}  & 
  \multicolumn{2}{c}{Scenario III}  & \multicolumn{2}{c}{Scenario IV}  \\ 
 &OM & \multicolumn{2}{c}{linear (1)} & \multicolumn{2}{c}{linear (1)} & 
      \multicolumn{2}{c}{non-linear (2)} & \multicolumn{2}{c}{non-linear (2)} \\ 
 &PM & \multicolumn{2}{c}{linear (1)} & \multicolumn{2}{c}{non-linear (2)} & 
     \multicolumn{2}{c}{linear (1)} & \multicolumn{2}{c}{non-linear (2)} \\ 
 Type & Estimator & CR & Length & CR & Length & CR & Length & CR & Length \\ 
  \hline
  \multicolumn{10}{c}{Continuous}\\
  \hline
  Basic & NN & 95.60 & 24.73 & 94.80 & 24.78 & 94.80 & 59.82 & 96.60 & 59.84 \\
        & GLM & 93.60 & 17.63 & 94.60 & 17.69 & 69.00 & 54.66 & 0.00 & 60.57 \\
        & DR (MLE) & 93.60 & 17.69 & 95.40 & 18.40 & 95.40 & 24.00 & 6.80 & 363.34 \\
        & DR (GEE) & 93.60 & 17.65 & 94.60 & 17.61 & 96.00 & 28.21 & 0.00 & 72.78 \\
  IPW MLE & IPW  & 95.80 & 22.36 & 10.40 & 122.99 & 99.20 & 38.23 & 2.40 & 787.33 \\ 
          & QBIPW1 & 94.80 & 21.67 & 98.20 & 47.09 & 100.00 & 45.92 & 83.40 & 92.19 \\
          & QBIPW2 & 95.60 & 23.55 & 97.59 & 55.69 & 99.60 & 56.49 & 96.98 & 102.93 \\
  IPW GEE & IPW  & 93.80 & 17.77 & 96.00 & 18.76 & 96.40 & 28.40 & 0.00 & 79.91 \\ 
          & QBIPW1 & 93.80 & 17.69 & 94.80 & 18.35 & 95.40 & 39.57 & 45.60 & 51.73 \\
          & QBIPW2 & 94.80 & 17.80 & 95.80 & 18.43 & 98.00 & 46.65 & 92.20 & 49.54 \\ 
 \hline
  \multicolumn{10}{c}{Binary}\\
  \hline
  Basic & NN & 94.20 & 3.82 & 95.00 & 3.82 & 93.20 & 5.34 & 95.00 & 5.34 \\ 
        & GLM & 94.80 & 1.36 & 94.60 & 1.35 & 95.00 & 2.11 & 0.20 & 1.91 \\
        & DR (MLE) & 92.80 & 1.30 & 94.00 & 1.32 & 95.80 & 1.72 & 0.00 & 1.97 \\ 
        & DR (GEE) & 92.60 & 1.30 & 94.00 & 1.31 & 94.80 & 1.71 & 0.00 & 1.77 \\ 
  IPW MLE & IPW  & 94.60 & 1.81 & 98.40 & 3.15 & 95.20 & 3.12 & 84.00 & 5.91 \\ 
          & QBIPW1 & 94.40 & 1.74 & 75.60 & 2.29 & 93.00 & 2.84 & 95.80 & 3.02 \\
          & QBIPW2 & 95.20 & 1.90 & 96.39 & 3.57 & 96.60 & 3.05 & 92.77 & 5.30 \\
  IPW GEE & IPW  & 93.20 & 1.46 & 0.00 & 0.95 & 94.80 & 2.43 & 18.00 & 1.22 \\
          & QBIPW1 & 94.20 & 1.43 & 49.60 & 1.08 & 96.20 & 2.37 & 96.40 & 2.24 \\
          & QBIPW2 & 94.20 & 1.44 & 93.80 & 1.07 & 96.80 & 2.40 & 96.40 & 2.36 \\ 
  \hline
\end{tabular}
\end{table}

In Scenario III, the proposed approaches yielded results that were close to the nominal rates which exhibited excellent performance. Finally, with respect to the Scenario IV, it can be observed that when both models are mis-specified, only those incorporating deciles and totals (along with the NN) in the case of continuous data and QBIPW1 (under MLE and GEE) for binary data, display a~coverage rate that is close to the nominal 95\%. In Section \ref{sec-quality} of the Appendix, we present a~detailed analysis of the quality of the reproduced totals and quantiles for the proposed methods. 

The simulation study illustrates the substantial advantages of integrating quantiles into IPW estimators. This approach markedly improves the robustness and accuracy of estimates, particularly in instances where both the outcome and probability models are mis-specified. QBIPW estimators demonstrate superior performance in complex, non-linear scenarios involving both continuous and binary outcomes, with GEE-based versions generally outperforming their MLE counterparts. It seems that the width of CI is comparable with the GLM and DR estimators, which suggests that using more constraints doesn't increase the variance. The proposed approach also provides narrower CI than the NN estimator. Consequently, researchers engaged in the analysis of continuous or count variables from multiple sources should consider employing quantile-based IPW methods to enhance the robustness of their results, especially when confronted with uncertain model specifications or complex relationships in real-world data analysis.

\section{Real data application}\label{sec-application}

\subsection{Data description}

In this section we present an attempt to integrate administrative and survey data about job vacancies for the end of 2022Q2 in Poland. The aim was to estimate the share of vacancies aimed at Ukrainian workers. After the Russian invasion of Ukraine on the 24th February 2022,  around 3.5 million persons (mainly women and children) arrived in Poland between 24th February and mid-May 2022 \citep{duszczyk2022war}. Some of them went to other European countries, but about 1 million stayed in Poland (as of 2023, cf. \citet{ukr-2023}).

The first source we used is the Job Vacancy Survey (JVS, known in Poland as the Labour Demand Survey), which is a~stratified sample of 100,000 units, with a~response rate of about 60\% ($S_{B}$). The survey population consists of companies and their local units with 1 or more employees. The sampling frame includes information about the industry code NACE (19 levels), region (16 levels), sector (2 levels), size (3 levels) and the number of employees according to administrative data integrated by Statistics Poland (RE). The JVS sample contains 304 strata created separately for enterprises with up to 9 employees and those with more than 10 employees \citep[cf.][]{jvs-meth}.

The survey measures whether an enterprise has a~job vacancy at the end of the quarter (on June 30, 2022) according to the following definition: \textit{vacancies are positions or jobs that are unoccupied as a~result of employee turnover or newly created positions or jobs that simultaneously meet the following three conditions}: (1) the jobs were actually vacant on the day of the survey, (2) the employer made efforts to find people willing to take on the job, (3) if suitable candidates were found to fill the vacancies, the employer would be willing to hire them. In addition, the JVS restricts the definition of vacancies by excluding traineeships, mandate contracts, contracts for specific work and business-to-business (B2B) contracts. Of the 60,000 responding units, around 7,000 reported at least one vacancy. Our target population included units with at least one vacancy, which according to the survey was between 38,000 and 43,000 at the end of 2022Q2.

The second source is the Central Job Offers Database (CBOP), which is a~register of all vacancies submitted to Public Employment Offices (PEOs -- $S_{A}$). CBOP is available online and can be accessed programmatically (via the API). CBOP includes all types of contracts and jobs outside Poland, so data cleaning was carried out to align the definition of a~vacancy with that used in the JVS. CBOP data collected via API include information about whether a~vacancy is outdated (e.g. 17\% of vacancies were outdated when downloaded at the end of  2022Q2). CBOP also contains information about unit identifiers (REGON and NIP), so we were able to link units to the sampling frame to obtain auxiliary variables with the same definitions as in the survey (24\% of records did not contain an identifier because the employer can withhold this information). The final CBOP dataset contained about 8,500 units included in the sampling frame. 

The overlap between JVS and CBOP was around 2,600 entities (around 4\% of the JVS sample and 30\% of CBOP), but only 30\% of these reported at least one vacancy in the JVS survey. This suggests significant under-reporting in the JVS, which, however, is a~problem beyond the scope of this paper. For the empirical study we treated both sources as separate and their correct treatment with the proposed methods will be investigated in the future.

\subsection{Analysis and results}

We defined our target variable as follows: $Y$ is \textit{the share of vacancies that have been translated into Ukrainian} (denoted as UKR), calculated separately for each unit. After the Russian aggression, many online job advertising services introduced information on whether the employer is particularly interested in employing Ukrainians (e.g. the Ukrainian version of the website was available, and job ads featured the following annotation: "We invite people from Ukraine").

\begin{table}[ht!]
\centering
\caption{Quantiles of registered employment (RE) for JVS, CBOP and the target variable UKR at the end of 2022Q2}
\label{tab-quants-data}
\begin{tabular}{rrrr}
  \hline
Decile of the RE & JVS & CBOP & CBOP-UKR \\ 
  \hline
  10\% & 3 & 4 & 4 \\ 
  20\% & 4 & 6 & 5 \\ 
  30\% & 5 & 8 & 8 \\ 
  40\% & 6 & 13 & 11 \\ 
  50\% & 8 & 20 & 18 \\ 
  60\% & 10 & 32 & 28 \\ 
  70\% & 17 & 51 & 48 \\ 
  80\% & 37 & 90 & 91 \\ 
  90\% & 100 & 211 & 215 \\ 
   \hline
\end{tabular}
\end{table}

Table \ref{tab-quants-data} shows the distribution of registered employment according to the JVS survey, for companies observed in the CBOP and companies willing to employ Ukrainians (with at least one vacancy translated into Ukrainian; denoted as CBOP-UKR). Companies observed in the CBOP are on average larger than those responding to the JVS. The median employment for JVS was 8, while for CBOP and CBOP-UKR -- 20 and 18 employees, respectively. This suggests that companies are selected into the CBOP depending on their size.

\begin{figure}[ht!]
    \centering
    \includegraphics[width=\textwidth]{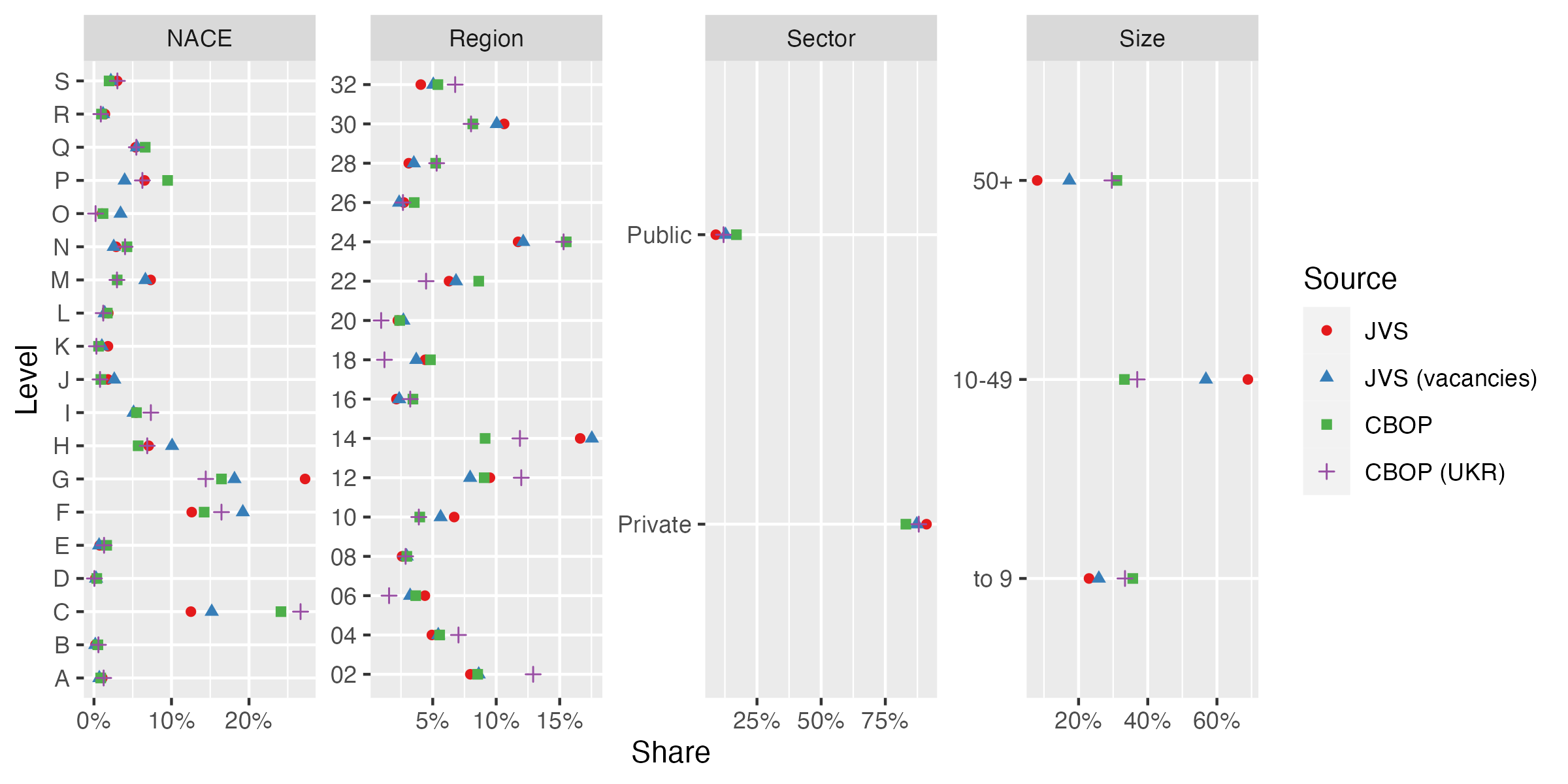}
    \caption{The distribution of 4 auxiliary variables in the sources at the end of 2022Q2. JVS: the total sample, JVS (vacancies): companies with at least one vacancy, CBOP: register data set, CBOP-UKR: units with at least one vacancy translated into Ukrainian}
    \label{fig-shares}
\end{figure}

The distributions of four categorical variables in the two sources are shown in Figure \ref{fig-shares}. The largest discrepancies between the sources can be seen in the case of company size, where the shares in CBOP are almost equal. There are some differences for selected regions (e.g. 14 Mazowieckie with Warsaw, the capital of Poland; 02 Dolnośląskie with Wrocław and 12 Małopolskie with Cracow). Regions from the eastern part of Poland (06 Lubelskie, 18 Podkarpackie, 20 Podlaskie), but also Pomorskie (22) with Tricity are characterised by the lowest share of vacancies for Ukrainians. The largest differences in terms of NACE are found for manufacturing (C), wholesale and retail trade (G) and hotels and restaurants (I).

\begin{figure}[ht!]
    \centering
    \includegraphics[width=\textwidth]{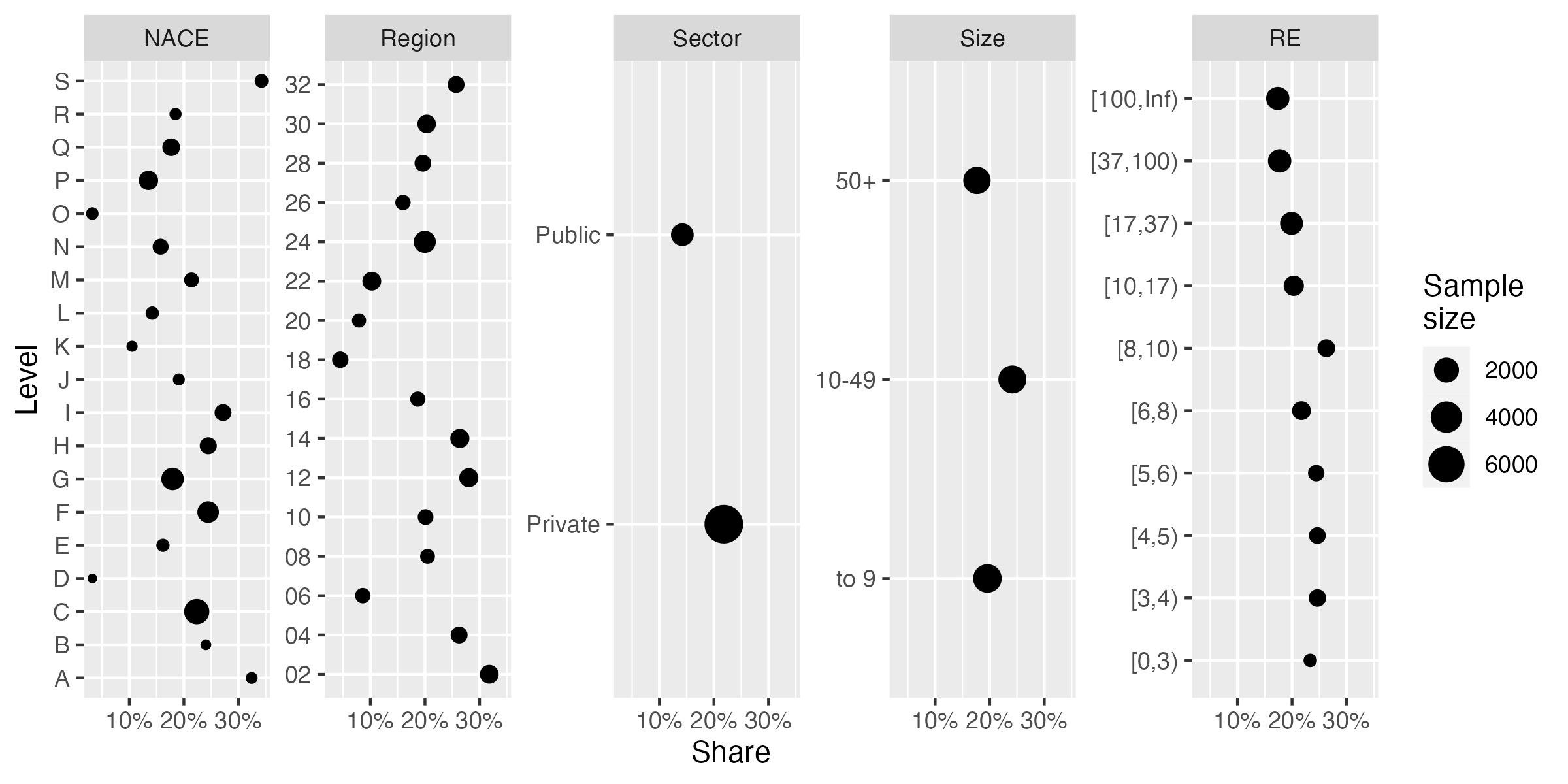}
    \caption{The target variable (the share of ads directed at Ukrainians) depending on 5 auxiliary variables at the end of 2022Q2}
    \label{fig-shares-ukr}
\end{figure}

Figure \ref{fig-shares-ukr} shows the share of vacancies targeted at Ukrainians according to four categorical variables and deciles of registered employment (RE) based on deciles estimated from the JVS survey. In general, the shares range between about 10\% and 30\%, especially across regions. The shares for the RE deciles range between 17\% for the largest units and 26\% for the median RE (8 employees). 

The following combinations of variables were considered: 

\begin{itemize}
    \item Set 0: Region (16 levels), NACE (19 levels), Sector (2 levels), Size (3 levels), $\log$(RE), $\log$(\#~vacancies) (the number of vacancies), $I$(\#~vacancies = 1) (whether employer seeks only  one person),
    \item Set 1A: Set 0 (without $\log$(RE)) + Quartiles of the RE (estimated from the JVS),
    \item Set 1B: Set 0 + Quartiles of the RE,
    \item Set 2A: Set 0 (without $\log$(RE)) + Deciles of the RE (without 10\%),
    \item Set 2B: Set 0 + Deciles of RE (without 10\%).
\end{itemize}

We decided not to include the number of vacancies for quantile balancing as almost 45\% of vacancies were equal to 1~for CBOP and almost 30\% for JVS. We consider the following estimators, assuming linear and logistic regression: GLM with Set 0, NN with set 0 (denoted as NN1), NN with $\log$(RE) and $\log$(\#~vacancies) only (denoted as NN2), DR GEE with Set 0, IPW estimators under MLE and GEE with the following sets: IPW with Set 0, IPW with Set 1A (QBIPW1A), IPW with Set 1B (QBIPW1B), IPW with Set 2A (QBIPW2A) and IPW with Set 2B (QBIPW1B).

Variance was estimated using the following bootstrap approach: 1) JVS sample was resampled using a~stratified bootstrap approach, 2) CBOP was resampled using simple random sampling with replacement. This procedure was repeated $B=500$ times. Table \ref{tab-results-emp} shows point estimates (denoted as Point), bootstrap standard errors (denoted as SE), the coefficients of variation (CV) and 95\% confidence intervals.

\begin{table}[ht!]
\centering
\caption{Estimates of the share of job vacancies aimed at Ukrainians at the end of 2022Q2 in Poland}
\label{tab-results-emp}
\begin{tabular}{lrrrrr}
  \hline
Estimator & Point & SE & CV & 2.5\% & 97.5\% \\ 
   \hline
  \multicolumn{6}{c}{Na\"ive }\\
  \hline
Na\"ive  & 20.51 & --  & -- & -- & -- \\ 
\hline
  \multicolumn{6}{c}{Estimators}\\
  \hline
  GLM & 22.68 & 0.59 & 2.61 & 21.52 & 23.84 \\
  NN1 & 25.88 & 6.41 & 24.78 & 13.31 & 38.45 \\
  NN2 & 23.38 & 1.91 & 8.18 & 19.63 & 27.13 \\
  DR GEE & 21.75 & 0.60 & 2.78 & 20.57 & 22.94 \\ 
  \hline
  \multicolumn{6}{c}{Inverse probability weighting (MLE)}\\
  \hline
  IPW & 22.88 & 0.80 & 3.52 & 21.31 & 24.46 \\  
  QBIPW1A & 22.24 & 0.76 & 3.43 & 20.74 & 23.73 \\ 
  QBIPW1B & 22.88 & 0.80 & 3.52 & 21.31 & 24.46 \\ 
  QBIPW2A & 22.26 & 0.78 & 3.48 & 20.75 & 23.78 \\ 
  QBIPW2B & 22.88 & 0.81 & 3.53 & 21.30 & 24.46 \\ 
  \hline
  \multicolumn{6}{c}{Inverse probability weighting (GEE)}\\
  \hline
  IPW & 21.75 & 0.60 & 2.78 & 20.57 & 22.94 \\ 
  QBIPW1A & 21.53 & 0.60 & 2.78 & 20.36 & 22.70 \\ 
  QBIPW1B & 21.69 & 0.60 & 2.78 & 20.51 & 22.87 \\ 
  QBIPW2A & 21.69 & 0.61 & 2.83 & 20.49 & 22.89 \\ 
  QBIPW2B & 21.74 & 0.61 & 2.83 & 20.53 & 22.94 \\ 
  \hline
\end{tabular}
\end{table}

The naive estimator suggests a~20.51\% share, while other methods provide slightly higher estimates. GLM and DR GEE estimators show similar results (22.68\% and 21.75\% respectively) with low standard errors. NN estimators show higher variability, with NN1 having the highest point estimate (25.88\%) but also the largest standard error. IPW and QBIPW estimators, both MLE and GEE-based, provide results ranging from 21.53\% to 22.88\%. Generally, GEE-based methods show lower standard errors and narrower confidence intervals compared to their MLE counterparts. Overall, most methods suggest the share of job vacancies for Ukrainians is between 21\% and 23\%, with GEE-based estimators showing the most precise estimates.

\section{Summary}\label{sec-summary}

Recent years have seen a~revolution in survey research with regard to the use and integration of new data sources for statistical inference. Efforts in this area have focused on the integration of survey data and inference based on non-probability samples, with applications in official statistics \citep{salvatore2023inference}. This is mainly because probability samples tend to be very costly and are associated with respondent burden. On the other hand, statistical inference based on non-probabilities samples is a~challenge, mainly owing to bias inherent in the estimation process. Consequently, it is necessary to develop new methods of statistical inference based on non-probability samples, which could be be useful in real applications \citep{wu2022statistical}. 

In this paper we have described quantile balancing inverse probability weighting for non-probability surveys. Based on our approach, for some of the methods considered in the article (IPW and DR) it is possible to reproduce not only totals for a~set of auxiliary variables but also for a~set of quantiles (or estimated quantiles). Such a~solution improves robustness against model mis-specification, helps to decrease bias and improves the efficiency of estimation. These gains have been confirmed by our simulation study, in which we have shown that the inclusion of quantiles for inverse probability weighting estimators improves the quality of estimates. 

The estimators proposed in this paper have the potential for real-life applications, provided that continuous variables of sufficient quality can be made available for use as auxiliary variables. The example of our empirical study, in which we estimated the share of job vacancies aimed at Ukrainian workers, demonstrates the effectiveness of the proposed methods in producing estimates of high precision.

Regarding aspects not addressed in this paper, one issue worth investigating further is the problem of under-coverage in non-probability samples, which is related to the violation of the positivity assumption (A2). The approach proposed in the paper equates the distributions of variables between the~probability (or the population) and the non-probability samples and can serve as an alternative to other approaches proposed in the literature, which consist, for example, of using an additional sample of individuals from the population not covered by the non-probability sample (Chen, Li \& Wu, \citeyear{chen_dealing_2023}).

\section*{Acknowledgements}

The authors' work has been financed by the National Science Centre in Poland, OPUS 20, grant no. 2020/39/B/HS4/00941. 

We thank reviewers and Associate Editor for the comments that improved the paper. In addition, authors would like to thank Łukasz Chrostowski who developed the \texttt{nonprobsvy} package which implements state-of-the-art estimators proposed in the literature. The package is available through \texttt{CRAN} and at \url{https://github.com/ncn-foreigners/nonprobsvy}. We also thank Natalie Shlomo and Tymon Świtalski for their valuable comments on the draft version of this paper.

Codes to reproduce the simulation study are freely available from the github repository: \url{https://github.com/ncn-foreigners/paper-nonprob-qcal}. An R package that implements joint calibration is available at \url{https://github.com/ncn-foreigners/jointCalib}. The package is based on calibration implemented in \texttt{survey}, \texttt{sampling} or \texttt{laeken} packages.

\printbibliography
\newpage
\appendix

\begin{center}
\begin{Large}
Appendix for the paper \textit{Quantile balancing inverse probability weighting for non-probability samples}
\end{Large}  
\end{center}
\allowdisplaybreaks

\section{Assumptions that guarantee unique solution to estimating equations}\label{appen-existance}

In the appendix we use notation $\bz_k = \left(\begin{smallmatrix}\bx_{k}\\\ba_{k}\end{smallmatrix}\right)$. 

\begin{itemize}
    \item[(B1)] The matrix $\displaystyle\sum_{k\in S_{A}}\bz_{k}\bz_{k}^{T}$ is positive definite.
    \item[(B2)] The matrix $\displaystyle\sum_{k\in S_{B}}\bz_{k}\bz_{k}^{T}$ is positive definite.
\end{itemize}

In the appendix we use the following notation:
\begin{align*}
    \pi(\bz; \bEta)  &:= 
    \frac{\exp\left(\bz^{T}\bEta\right)}{1+\exp\left(\bz^{T}\bEta\right)},\\
    \pi(t)  &:= \frac{\exp\left(t\right)}{1+\exp\left(t\right)},\\
    \pi'(\bz; \bEta) &:= \pi(\bz; \bEta)\left(1-\pi(\bz; \bEta)\right).
\end{align*}

\begin{tw}
    Assumption (B1) is a necessary and sufficient condition for existence of a unique solution to: 
    \begin{equation*}
        \bG(\bEta) = 
        \sum_{k \in S_{A}} \frac{\bz_{k}}{\pi\left(\bz_{k}; \bEta\right)} -
        \sum_{k \in S_{B}} d_{k}^{B} \bz_{k} = \boldsymbol{0},
    \end{equation*}
    and assumption (B2) is a necessary and sufficient condition for existence of a unique solution to: 
    \begin{equation*}
        \bU(\bEta)=\sum_{k \in S_{A}} \bz_{k}-\sum_{k \in S_{B}} d_{k}^{B}
        \pi\left(\bz_{k}; \bEta\right) \bz_{k}=\boldsymbol{0}.
    \end{equation*}
\end{tw}

Conditions (B1) and (B2) are easy to check through the process of data analysis. If either of these is violated a potential remedy is to reduce the number of quantiles used. Notice that (B1) implies $n_{A}\geq J$ where $J$ is the dimension of $\bz$ and that there are at least $J$ linearly independent $\bz$'s in $S_{A}$. (B2) implies analogous properties of $S_{B}$.

\begin{proof}
    We will first prove that $\ell^{\ast}$ (the anti-derivative of $\bU$) has a unique global maximum. Notice that if (B2) holds the hessian of $\ell^{\ast}$:
    \begin{equation*}
        \frac{\partial\ell^{\ast}}{\partial\bEta\partial\bEta^{T}}(\bEta)=\frac{\partial}{\partial\bEta}\bU(\bEta)=-\sum_{k\in S_{B}}d_{k}^{B}
        \pi'(\bz; \bEta)\bz_{k}\bz_{k}^{T},
    \end{equation*}
    is positive definite for any $\bEta\in\mathbb{R}^{J}$ so $\ell^{\ast}$ is strictly concave. Therefore, it has at most one global maximum and since hessian is always invertible the second derivative test tells us that $\bU(\bEta)$ has at most one root.
    
    Now notice that (B2) implies, by rank-nullity theorem, that we have:
    \begin{equation*}
        \text{Span}\left(\bz_{1}, \ldots, \bz_{n_{A}}\right)^{\perp} =
        \left(\mathbb{R}^{J}\right)^{\perp} =
        \{\boldsymbol{0}\},
    \end{equation*}
    where $(\cdot)^{\perp}$ denotes the orthogonal component, which in turn gives us:
    \begin{equation}\label{asdsfasdfdafdsafda}
        \exists j\in S_{A}:|\bz_{j}^{T}\bEta_{m}|\rightarrow\infty,
    \end{equation}
    so $\pi_{j}^{A}$ tends to either $0$ or $1$. This implies that the PL function:
    \begin{align*}
        \ell^{*}(\bEta)&=\sum_{k \in S_{A}} \left(\log \left(\pi_{k}^{A}\right)-
        \log\left(1-\pi_{k}^{A}\right)\right)+
        \sum_{k \in S_{B}} d_{k}^{B} \log \left(1-\pi_{k}^{A}\right),
    \end{align*}
    which is bounded from above must tend to $-\infty$.     

    Similarly an analogous reasoning for any anti-derivative of $\bG$, for example:
    \begin{align*}
        H(\bEta) &= \sum_{k\in S_{A}}\int_{q}^{\bEta^{T}\bz_{k}}
        \frac{1}{\pi(t)}dt-
        \sum_{k\in S_{B}}d_{k}^{B}\bEta^{T}\bz_{k},
    \end{align*}
    for some $q\in\mathbb{R}$, the hessian of which is:
    \begin{equation*}
        \frac{\partial}{\partial\bEta}\bG(\bEta)=
        -\sum_{k\in S_{A}}\frac{\bz_{k}\bz_{k}^{T}}{\pi(\bz_{k};\bEta)^{2}}
        \pi'(\bz_{k};\bEta),
    \end{equation*}
    guarantees that $\bG$ has at most one root. For reasons that will become clearer down the line we choose $q<0$ such that:
    \begin{equation*}
        \left\lVert\sum_{k\in S_{B}}d^{B}_{k}\bz_{k}-\sum_{k\in S_{A}}\bz_{k}\right\rVert<n_{A}\left(e^{-q}-q\right),
    \end{equation*}
    which is possible since $q\mapsto e^{-q}-q$ is a continuous bijection from $\mathbb{R}$ to $\mathbb{R}$.
    
    Strict concavity of $H$ implies that a sufficient condition for existence of a root of $\bG$ is:
    \begin{equation}\label{limit-to-proof}
        \lim_{\lVert\bEta\rVert\rightarrow\infty}H(\bEta)=-\infty,
    \end{equation}
    where $\lVert\cdot\rVert$ denotes the euclidean norm $\boldsymbol{\zeta}\overset{\lVert\cdot\rVert}{\mapsto}\sqrt{\left|\boldsymbol{\zeta}^{T}\boldsymbol{\zeta}\right|}$.
    To prove \eqref{limit-to-proof} it is sufficient to prove that $\displaystyle\lim_{m\rightarrow\infty}|H(\bEta_{m})|=\infty$ for sequences $(\bEta_{m})_{m\in\mathbb{N}}\subseteq\mathbb{R}^{J}$ with $\displaystyle\lim_{m\rightarrow\infty}\lVert\bEta_{m}\rVert=\infty$, such that the sequence $\text{sgn}(\bEta_{m})=\frac{\bEta_{m}}{\lVert\bEta_{m}\rVert}$ is constant.

    To see that the claim above is true first observe that for any vector $\bEta$ and $s\geq1$:
    \begin{align*}
        H(s\bEta)-sH(\bEta)&=
        \sum_{k\in S_{A}}\left(\int_{q}^{s\bEta^{T}\bz_{k}}\frac{1}{\pi(t)}dt-
        \int_{q}^{\bEta^{T}\bz_{k}}\frac{s}{\pi(t)}dt\right)\\
        &\leq
        \sum_{k\in S_{A}}\left(\int_{q}^{\bEta^{T}\bz_{k}}\frac{s}{\pi(st)}dt+
        \int_{q/s}^{q}\frac{s}{\pi(st)}dt-
        \int_{q}^{\bEta^{T}\bz_{k}}\frac{s}{\pi(t)}dt\right)\\
        &=\sum_{k\in S_{A}}\int_{q}^{\bEta^{T}\bz_{k}}
        \frac{e^{-s}-1}{e^{t}}dt+
        n_{A}\int_{q/s}^{q}\frac{s}{\pi(st)}dt\\
        &=\sum_{k\in S_{A}}\left(e^{-s}-1\right)
        \left(-e^{-\bEta^{T}\bz_{k}}+e^{-q}\right)+
        n_{A}\int_{q/s}^{q}\frac{s}{\pi(st)}dt\\
        &=\left(1-e^{-s}\right)
        \left(\sum_{k\in S_{A}}e^{-\bEta^{T}\bz_{k}}-n_{A}e^{-q}\right)+
        n_{A}\left(sq-q+e^{-q}-e^{-qs}
        \right),
    \end{align*}
    where the second line follows by substitution $us=t, sdu=dt$, and the last term is bounded from above in $s$. By local boundedness of $H$ there exists a non-negative constant $R$ such that:
    \begin{equation*}
        \forall s>0: H\left(s\bEta\right)\leq
        sH\left(\bEta\right)+R.
    \end{equation*}
    Thus, we have for any sequence of normed vectors $\left(\boldsymbol{\omega}_{m}\right)_{m\in\mathbb{N}}\subseteq\mathbb{R}^{J}$:
    \begin{align*}
        \sup\{t>0: H(t\boldsymbol{\omega}_{m})>-M\}&\leq
        \sup\{t>0: tH(\boldsymbol{\omega}_{m})>-M-R\}\\
        &\leq\frac{-M-R}{H(\boldsymbol{\omega}_{m})}\rightarrow\infty
        \Longleftrightarrow H(\boldsymbol{\omega}_{m})\rightarrow0.
    \end{align*}
    However, by continuity the last implies that for every $\varepsilon>0$ there exists $\boldsymbol{u}$ with $\lVert\boldsymbol{u}\rVert=1$ and $H(\boldsymbol{u})>-\varepsilon$, but this is impossible since then:
    \begin{align*}
        \sum_{k\in S_{B}}d_{k}^{B}\boldsymbol{u}^{T}\bz_{k}-\varepsilon&>
        \sum_{k\in S_{A}}\int_{q}^{\boldsymbol{u}^{T}\bz_{k}}\frac{1}{\pi(t)}dt=
        n_{A}(e^{-q}-q)+\sum_{k\in S_{A}}
        \left(\boldsymbol{u}^{T}\bz_{k}+e^{-\boldsymbol{u}^{T}\bz_{k}}\right)\\
        \implies
        \left\lVert\sum_{k\in S_{B}}d_{k}^{B}\bz_{k}-
        \sum_{k\in S_{A}}\boldsymbol{u}^{T}\bz_{k}\right\rVert
        &\geq\left|\boldsymbol{u}^{T}\left(\sum_{k\in S_{B}}d_{k}^{B}\bz_{k}-
        \sum_{k\in S_{A}}\boldsymbol{u}^{T}\bz_{k}\right)\right|\\
        &\geq n_{A}(e^{-q}-q)+\sum_{k\in S_{A}}e^{-\boldsymbol{u}^{T}\bz_{k}}+\varepsilon,
    \end{align*}
    which is impossible by virtue of how constant $q$ was chosen. Therefore we have that:
    \begin{equation*}
        \forall M>0:
        \sup_{\lVert \boldsymbol{\omega}\rVert=1}
        \inf\{t>0: \forall s> t: H(s\boldsymbol{\omega})<-M\}=
        \sup_{\lVert \boldsymbol{\omega}\rVert=1}
        \sup\{t>0: H(t\boldsymbol{\omega})\geq-M\}<\infty.
    \end{equation*}
    Now assume that:
    \begin{equation*}
        \forall M>0, \lVert \boldsymbol{\omega}\rVert=1\enskip
        \exists N>0\enskip\forall r>N: H(r\boldsymbol{\omega})<-M.
    \end{equation*}
    Choose any $M>0$ and let:
    \begin{equation*}
        T := \sup_{\lVert \boldsymbol{\omega}\rVert=1}
        \sup\{t>0: H(t\boldsymbol{\omega})\geq-M\}<\infty.
    \end{equation*}
    Now for any $r>T$ we have:
    \begin{equation*}
        \forall \lVert z\rVert=1:H(r\boldsymbol{z})<-M.
    \end{equation*}
    Putting it altogether yields:
    \begin{equation*}
       \forall M>0 \enskip\exists T>0 \enskip\forall r>T \enskip
       \forall \lVert \boldsymbol{\omega} \rVert=1:
       H(r\boldsymbol{z})<-M.
    \end{equation*}
    It is sufficient to check that:
    \begin{equation*}
        H(\bEta_{m})\rightarrow\infty,
    \end{equation*}
    for sequences of $\bEta_{m}$ with constant sign and increasing to infinity in norm as was claimed previously.

    Now we proceed to prove the central claim about $H$. We have already proved that:
    \begin{equation*}
        \forall\bEta\in\mathbb{R}^{J}\enskip\exists R>0\enskip\forall s>0: 
        H\left(s\bEta\right)\leq sH\left(\bEta\right)+R,
    \end{equation*}
    and chosen $q$ such that:
    \begin{equation*}
        \forall\lVert\boldsymbol{z}\rVert=1:H(\boldsymbol{z})<0.
    \end{equation*}
    The convergence to $-\infty$ for sequences $\left(\bEta_{m}\right)_{m\in\mathbb{N}}$ defined previously is straightforwardly attained.
    
    It is not difficult to see that (B2) is in fact necessary for the existence of a unique solution to $\bG=\boldsymbol{0}$. If (B2) is violated then, since the matrix in (B2) is clearly non-negative define, we can pick some $\boldsymbol{\xi}$ such that:
    \begin{equation*}
        \boldsymbol{0}\neq\boldsymbol{\xi}\in
        \text{Span}\left(\bz_1, \ldots, \bz_{n_{A}}\right)^{\perp}.
    \end{equation*}
    Now if $\bG=\boldsymbol{0}$ has a solution $\bEta$ then:
    \begin{equation*}
        \boldsymbol{0} = \bG(\bEta) = 
        \sum_{k \in S_{A}} \frac{\bz_{k}}{\pi\left(\bEta^{T}\bz_{k}\right)} -
        \sum_{k \in S_{B}} d_{k}^{B} \bz_{k} =
        \sum_{k \in S_{A}} \frac{\bz_{k}}{\pi\left(\bEta^{T}\bz_{k} + \boldsymbol{\xi}^{T}\bz_{k}\right)} -
        \sum_{k \in S_{B}} d_{k}^{B} \bz_{k} = \bG(\bEta + \boldsymbol{\xi}).
    \end{equation*}
    So $\boldsymbol{\xi}+\bEta$ is also a solution to $\bG=\boldsymbol{0}$ and therefore the space of solutions to the GEE equation has exactly dimension equal to $\displaystyle J-\text{Rank}\left(\sum_{k\in S_{A}}\bz_{k}\bz_{k}^{T}\right)$ if such a solution exists. An analogous arguments shows that assumption (B1) is not only sufficient but also necessary for the existence of a unique solution to $\bU=\boldsymbol{0}$.
\end{proof}

\clearpage
\section{Quality of reproduced totals and quantiles}\label{sec-quality}

This section presents the results of the quality assessment of the total and quantile reproduction for all IPW estimators employed in the simulation study. For the following quantities and each simulation run ($r=1,\ldots,500)$ the euclidean norm of the difference between them and their estimates respectively have been calculated as:

\begin{itemize}
    \item Population size:
    $$
    \nu_{r,\hat{N}} = |\hat{N}_r- N|.
    $$
      \item Quantiles (for $x_1$ and $x_2$ jointly, considering both quartiles 
      and deciles, \\ where $\alpha \in \{0.25,0.50,0.75, 0.1, ..., 0.9\}$): 
    $$
    \nu_{r, \hat{Q}} = \sqrt{\sum_{\alpha}\sum_{p=1}^2\left(\hat{Q}_{r, x_p,\alpha}- Q_{r, x_p,\alpha}\right)^2}.
    $$
    \item Totals (for $x_1$ and $x_2$ jointly): 
    $$
    \nu_{r, \hat{\tau}}  = \sqrt{\sum_{p=1}^2\left(\hat{\bar{\tau}}_{r,x_p}- \tau_{r,x_p}\right)^2}.
    $$
\end{itemize}

\noindent The reference values for $\hat{\tau}_{x_p,r}$ and $Q_{r, x_p,\alpha}$ are estimated from the probability sample $S_B$. These may vary in the simulation as we do not calibrate the $S_B$ sample.

Table \ref{tab-quality} reports both the mean and median values of $\nu_{r,\hat{N}}$, $\nu_{r, \hat{\tau}}$, and $\nu_{r, \hat{Q}}$. The inclusion of both measures is intended to account for instances where the algorithm did not converge, resulting in imperfect reproduction of totals and quantiles for the GEE estimator.

\begin{table}[ht!]
\centering
\small
\caption{Quality of reproduced population size, totals and quantiles}
\label{tab-quality}
\begin{tabular}{lrrrrrr}
  \hline
  Estimator & \multicolumn{3}{c}{PM1} & \multicolumn{3}{c}{PM2} \\ 
     & $\nu_{r,\hat{N}}$ & $\nu_{r, \hat{Q}}$ & $\nu_{r, \hat{\tau}}$ & $\nu_{r,\hat{N}}$ & $\nu_{r, \hat{Q}}$ & $\nu_{r, \hat{\tau}}$  \\
     \hline
\multicolumn{7}{c}{Mean from simulations} \\
\hline
  Naive & -- & 0.52 & -- & -- & 2.28 & -- \\ 
  \hline
  \multicolumn{7}{c}{MLE} \\
  \hline
  IPW & 658.81 & 0.16 & 1049.47 & 11779.12 & 2.98 & 78050.24 \\ 
  QBIPW1 & 726.83 & 0.16 & 1094.00 & 3403.08 & 0.90 & 13198.45 \\ 
  QBIPW2 & 1334.84 & 0.15 & 1884.13 & 7310.56 & 0.46 & 20054.55 \\ 
  \hline
  \multicolumn{7}{c}{GEE} \\
  \hline
  IPW & 0.00 & 0.15 & 0.00 & 0.00 & 1.37 & 0.00 \\ 
  QBIPW1 & 0.11 & 0.10 & 2.74 & 0.00 & 0.36 & 0.00 \\  
  QBIPW2 & 133.95 & 0.05 & 406.16 & 41.10 & 0.05 & 36.12 \\ 
  \hline
\multicolumn{7}{c}{Medians from simulations} \\
\hline
Naive & -- & 0.51 & -- & -- & 2.28 & -- \\ 
\hline
\multicolumn{7}{c}{MLE} \\
\hline
  IPW & 358.00 & 0.15 & 760.42 & 10617.02 & 2.76 & 70068.24 \\ 
  QBIPW1 & 353.72 & 0.15 & 732.87 & 3081.35 & 0.86 & 10515.02 \\ 
  QBIPW2 & 442.59 & 0.12 & 901.13 & 3242.52 & 0.41 & 8157.91 \\ 
  \hline
  \multicolumn{7}{c}{GEE} \\
  \hline
  IPW & 0.00 & 0.14 & 0.00 & 0.00 & 1.37 & 0.00 \\ 
  QBIPW1 & 0.00 & 0.10 & 0.00 & 0.00 & 0.36 & 0.00 \\ 
  QBIPW2 & 0.00 & 0.04 & 0.00 & 0.00 & 0.05 & 0.00 \\ 
   \hline
\end{tabular}
\end{table}

As anticipated based on the design, GEE-based estimators demonstrate superior performance compared to MLE-based methods across all measures. It is noteworthy that GEE-based methods demonstrate superior performance with regard to quantile reproduction, with QBIPW2 achieving the lowest error rates. The proposed QBIPW1 and QBIPW2 methods demonstrate exceptional performance, particularly in the context of the more challenging PM2 scenario. This is due to their design, which accounts for both quantiles and totals. QBIPW1 accounts for quartiles, while QBIPW2 utilises deciles. In the interpretation we focus on the \textit{median} results.

The results demonstrate a~pronounced discrepancy in performance between the PM1 and PM2 scenarios, with all estimators exhibiting better outcomes under PM1. This discrepancy is apparent not only in the reproduction of population size and totals, but also in the reproduction of quantiles. For example, under PM1, the majority of methods achieve mean quantile errors below 0.2, whereas under PM2, these errors frequently exceed 1.0 for MLE-based methods. Nevertheless, the proposed GEE-based QBIPW1 and QBIPW2 methods demonstrate remarkable resilience in quantile estimation, even under PM2 (with mean quantile errors as low as 0.05 for QBIPW2), thereby illustrating their capacity to capture the population's distribution across diverse probability models. The consistent performance in quantile reproduction, coupled with the accurate estimation of population size and totals, serves to reinforce the efficacy of the proposed methods in incorporating both quantile and total information in IPW estimation. The superior performance of QBIPW1 and QBIPW2 under PM2 serves to demonstrate their capacity to handle complex probability models by leveraging a more comprehensive set of population characteristics.

\end{document}